\documentclass[12pt,draftclsnofoot,onecolumn]{IEEEtran}

\usepackage{algorithm} 
\usepackage{algorithmic}  
\usepackage[algo2e]{algorithm2e}

\renewcommand{\algorithmcfname}{ALGORITHM}
\SetAlFnt{\small}
\SetAlCapFnt{\small}
\SetAlCapNameFnt{\small}
\SetAlCapHSkip{0pt}
\IncMargin{-\parindent}
\usepackage{algorithm}
\usepackage{moreverb,algorithmic,subfigure}
\usepackage{amsmath}
\usepackage{amssymb,lscape}
\usepackage{tabularx}
\usepackage{balance}
\usepackage{multirow}
\usepackage{comment}

\newcommand{\be}{\begin{equation}}
\newcommand{\ee}{\end{equation}}
\usepackage{cleveref}
\usepackage{cite}
\usepackage{epsfig}
\usepackage{times}
\usepackage{graphicx}
\usepackage[figuresright]{rotating}
\usepackage{amssymb}
\usepackage{amsmath}
\usepackage{setspace}
\usepackage{rotating}
\usepackage{multirow}
\usepackage{wrapfig}
\usepackage{booktabs}
\usepackage{array}
\usepackage{comment}
\usepackage{subfigure}
\usepackage{epstopdf}
\usepackage{epsfig}
\usepackage{amsmath}
\usepackage{amssymb}
\usepackage{multicol}
\usepackage{caption}
\usepackage{times}

\usepackage{moreverb,algorithmic,subfigure}
\usepackage{amsmath}
\usepackage{amssymb,lscape}
\usepackage{tabularx}
\usepackage{times}
\usepackage{float}

\usepackage{caption}
\usepackage{varwidth,xcolor}
\usepackage[margin=2cm]{geometry}

\providecommand{\keywords}[1]{\textbf{\textit{Index terms---}} #1}
\usepackage{colortbl}

\begin{document}
\title{\LARGE{Configuration Learning in Underwater Optical Links}}
\author{Xueyuan Zhao, Zhuoran Qi and Dario Pompili\\
\thanks{The authors are with the Department of Electrical and Computer Engineering~(ECE), Rutgers University--New Brusnwick, NJ, USA. E-mails: \{xueyuan.zhao, zhuoran.qi, pompili\}@rutgers.edu.}
\thanks{This work was supported by the National Science Foundation~(NSF) NeTS and CPS Awards, No.~1763964 and 1739315, respectively.}
}


\clearpage
\maketitle
\thispagestyle{empty}\pagenumbering{arabic}


\begin{abstract}
A new research problem named configuration learning is described in this work. A novel algorithm is proposed to address the configuration learning problem. The configuration learning problem is defined to be the optimization of the Machine Learning~(ML) classifier to maximize the ML performance metric optimizing the transmitter configuration in the signal processing/communication systems. Specifically, this configuration learning problem is investigated in an underwater optical communication system with signal processing performance metric of the physical-layer communication throughput. A novel algorithm is proposed to perform the configuration learning by alternating optimization of key design parameters and switching between several Recurrent Neural Network~(RNN) classifiers dependant on the learning objective. The proposed ML algorithm is validated with the datasets of an underwater optical communication system and is compared with competing ML algorithms. Performance results indicate that the proposal outperforms the competing algorithms for binary and multi-class configuration learning in underwater optical communication datasets. The proposed configuration learning framework can be further investigated and applied to a broad range of topics in signal processing and communications.









\end{abstract}


\section{Introduction}
Underwater wireless communications have the potential to transform how underwater explorations are performed. Application areas include underwater natural resources detection, submarines, and underwater marine science, to name just a few. Nowadays, mainstream underwater wireless communications are based on underwater acoustic signals and underwater wireless optical signals. For underwater acoustic communications, although the communication range can be tens of kilometers, the data rate is the major bottleneck, with maximum values of a few hundred Kbps~\cite{Stojanovic2006}. In contrast, the data rate of an underwater optical communication system can be in the Gbps scale. Compared with the underwater acoustic system of approximate $1.5~\rm{km/s}$ of propagation speed and maximum a few $\rm{kHz}$ of bandwidth, the underwater wireless optical system has approximate $2.26 \times 10^{5}~\mathrm{km/s}$ propagation speed and $\rm{MHz}$ scale bandwidth, resulting in very low latency and much higher data rate compared with acoustic communication systems. However, the coverage becomes the major constraint~\cite{Doniec13,Kaushal16} since optical signals attenuate significantly in the underwater channel, resulting in short communication distances from a few to tens of meters. To improve the data rate of underwater acoustic communication systems, we have proposed Acoustic Massive Multi-Input Multi-Output and Carrier Aggregation (AMMCA) techniques~\cite{AMMCA15}. We have further investigated the proposed underwater acoustic carrier aggregation technique by simulations~\cite{Zhao16} and real ocean experiments~\cite{Zhao17}. There are research works~\cite{Demirors16,Lawal16} in other research groups working on similar proposals after our work on AMMCA. Expanded bandwidth is designed~\cite{Demirors16} in an acoustic hardware system, and the data rate improvements are tested and verified by experiments. Multi-Input Multi-Output~(MIMO) capacity has been analyzed for a 2-by-2 system~\cite{Dong15} and for downlink underwater optical systems~\cite{Zhang15}. MIMO has been proposed for the underwater optical system and the error rate performance has been analyzed~\cite{Jamali17}. Impulse response modeling of the optical MIMO channel has been studied~\cite{Zhang16}. Experimental works have shown that an Orthogonal Frequency Division Multiplexing~(OFDM)-based laser communication system with a maximum optical power of $30~\rm{mW}$, modulated by $10~\rm{Gbps}$ 16-QAM data signal with $5~\rm{GHz}$ sampling rate, has achieved $8.8~\rm{Gbps}$ transmission data rate~\cite{Ho17} but under very short, $6.4~\rm{m}$ distance in the water tank. Similar works~\cite{Xu16,Lu16} on OFDM-based underwater laser communication have shown the system with power up to $30~\rm{mW}$ to achieve the $~\rm{Gbps}$ scale communication data rate with a very short distance of less than $10~\rm{m}$ in the tank. In this work, the underwater optical communication system is adopted as the communication system under study to evaluate the proposed configuration learning framework.

\textbf{Existing Works:}
Machine learning and deep learning are among the most promising directions of signal processing in communication system research nowadays and are undergoing significant research activities producing a large number of novel outputs~\cite{Jagannath19, Wang19}. The existing works on machine learning and deep learning in physical-layer signal processing research cover on major research problems including signal detection~\cite{OShea17, Wang17, Shlezinger19, Farsad18, Samuel17}, channel estimation~\cite{He18, Ye18, Huang18}, Channel State Information~(CSI) compression~\cite{Wen18}, precoding~\cite{Huang19}, and interference mitigation~\cite{He17}. On signal detection, the autoencoder structures are described for data symbol detection~\cite{OShea17} as well as decoding~\cite{Wang17}. The revised Viterbi decoding structure without requiring CSI is proposed in~\cite{Shlezinger19} by replacing the CSI-based computation with Deep Neural Network~(DNN). A Sliding Bidirectional Recurrent Neural Network~(SBRNN) is proposed in~\cite{Farsad18} to detect the data symbols. A specific deep neural network structure is theoretically derived based on zero-forcing detection~\cite{Samuel17}. On channel estimation, DNN is studied for channel estimation and compared with Least Square~(LS) and Minimum Mean Square Error~(MMSE) algorithms~\cite{Ye18}. The structure of Learned Denoising-based Approximate Message Passing~(LDAMP) network is applied to the channel estimation in millimeter-wave massive MIMO systems~\cite{He18}. The DNN is investigated on the channel estimation and Direction-of-Arrival~(DoA) estimation in massive MIMO systems~\cite{Huang18}. There are existing works on machine learning/deep learning applied to physical-layer link adaptation. An autoencoder-based method is studied to extract feature from estimated Channel State Information~(CSI) to determine modulation and coding scheme~\cite{Dong18}. The multilayer feedforward neural network is studied on the modulation and coding scheme adaptation with features from estimated CSI and Signal-to-Noise Ratio~(SNR)~\cite{Tato19}. The Link Adaptation~(LA) by machine learning in 802.11 vehicular network is discussed in~\cite{Xu19} with the assumption of available CSI information.
{
This work distinguishes from the above existing works in two ways. First, the configuration learning problem is a new research problem. The configuration learning is defined by the optimization of the ML classifier to adapt the physical-layer communication configuration. The focus of the concept configuration learning is on the optimization of the ML classifier performance. In contrast, the existing rate adaptation works are focusing on the optimization of the physical-layer performance. Therefore, our work is different by the concept from the existing works on machine learning in link adaptation. The differences between the configuration learning and rate adaptation are further explained in mathematical form in Sect.~\ref{sec:problem}. Second, an original ML algorithm is proposed to optimize the classifier to address the configuration learning problem in the underwater optical communication systems. In this new ML algorithm, there is alternating optimization to tune the key parameters in several Recurrent Neural Network~(RNN)-based classifiers, and the switching of the RNN classifier according to the configuration classification tasks. This new algorithm, named by \emph{SwitchOpt~RNN}, outperforms the competing ML classifiers. The details of the proposed algorithm are presented in Algorithm~\ref{alg:proposal_config_learning}. With these two major innovations, our work has a high level of novelty compared with all existing works on machine learning/deep learning in the signal processing/communication physical-layer research.

To validate our configuration learning proposal and the new algorithm, the datasets are generated from a physical-layer simulator by extracting the received signal and recoding the labels of the transmitter configuration maximizing the physical-layer throughput. These datasets contain the key information on the channel as well as the signal characteristics generated by the transmitter. The physical-layer simulator adopted in the dataset generation has the underwater optical channel models and receiver signal processing chains implemented. The generated datasets are applied to the training of the ML classifiers and the validation. Our proposal is compared with the competing ML classifiers including deep-learning-based algorithms of LSTM, bi-directional LSTM~(Bi-LSTM) and Gated Recurrent Unit~(GRU), decision tree, Adaptive Boosting ensemble~(AdaBoost), and Support Vector Machine~(SVM).

The proposal of this new algorithm is motivated by several key observations: 1)~RNN-based classifiers consistently outperform the ML classifiers of the decision tree, AdaBoost, and SVM. This performance gain is because the RNN-based structure has deep memory that can learn the features in the sequential signal and separate subtle differences in the sequential signal. The signal received at the receiver of the underwater optical communication system is the sequential signal mostly suited to be classified by the RNN-based ML classifiers. 2)~There are optimal key parameters including the number of hidden units and the number of epochs, in tuning the performance of several RNN classifiers of LSTM, Bi-LSTM, and GRU. The joint optimization is needed to tune these parameters so as to improve the training and optimization efficiency, In this work, we are proposing the alternating optimization to address this joint optimization problem. 3)~The optimal RNN classifier is dependent on the objective of the configuration learning. In the underwater communication system, we have tested several learning objectives, including varying the coding rate of the Turbo code, and the spreading factor. The ML classifier designed should be able to adapt to the learning objective of the configuration learning. Therefore, we further propose the switching of the RNN classifier according to the training results of the configuration learning process. All these three design considerations have produced our innovation algorithm, the \emph{SwitchOpt RNN} proposal.
}

\textbf{Our Contributions:}
The major contributions of this work are summarized as follows.
{
\begin{enumerate}

\item The configuration learning research problem is defined to be the problem of optimizing the ML classifier for the configuration classification in signal processing and communication systems. Specifically, the configuration learning problem is investigated on the transmitter configuration learning in an underwater optical communication system.

\item A configuration learning algorithm is proposed for the underwater optical communication system under study. This algorithm is based on the alternating optimization of the RNN-based ML classifier together with the switching of the RNN classifier based on the learning objective. This algorithm is named \emph{SwitchOpt~RNN} by the techniques adopted in the design. The performance of this algorithm is compared with several algorithms including LSTM, Bi-LSTM, GRU, decision tree, AdaBoost, and SVM. The associated training dataset generation method is proposed for the ML classifier evaluated. This dataset contains the received signal waveforms as well as the labels of the optimal transmitter configuration maximizing the physical-layer throughput. 

\item The results indicate that the proposed algorithm, \emph{SwitchOpt~RNN}, outperforms the competing ML algorithms in the configuration learning problem of the underwater optical communication system for all the three binary classification cases and two multi-class classification cases. The proposed algorithm is therefore suited for the configuration learning to adapt the coding rate and spreading factor of the underwater optical system. 

\item An OFDM-based optical communication structure is proposed with the transmitter and receiver signal processing chains for the evaluation of the proposed configuration learning method. the phase distortion caused by underwater scattering effects is handled by conjugate-symmetric OFDM modulation at the transmitter and the pilot-based channel estimation and signal detection at the receiver. The Inter-Symbol-Interference~(ISI) by optical multipath is mitigated by the OFDM modulation.

\end{enumerate}

Regarding the verification of the proposed framework and the algorithm, the datasets are generated by the physical-layer simulator that emulate the physical-layer throughput performance with Turbo code and all components of the signal processing chain, including practically implemented channel estimator and signal detector that are very close to the hardware implementation. Furthermore, the channel simulation in the physical-layer simulator adopts channel models and parameters verified by field experiments. Therefore, the results obtained are expected to represent the experiment results closely. In addition, the configuration learning problem has the assumption that neither SNR nor CSI is needed to perform the transmitter configuration learning. Only received signal waveform is needed for the trained ML classifier to identify the optimal transmitter configuration to achieve the high classification accuracy. Therefore, our approach is well suited for the systems with no interface to acquire SNR or CSI information at the receiver.
}

As far as the design of the underwater optical system is concerned, there is an issue regarding the phase distortion by the optical signal dispersion in the underwater optical channel. It is necessary to design a phase-robust modulation scheme in the optical transmitter. The approach adopted in this work is to design an intensity-modulation scheme where the information is only encoded by the intensity of the optical signal. This modulation is realized by a particular OFDM modulation where the information symbols are placed in a conjugate-symmetric way in the OFDM modulation. Only one stream signal is sent at the transmitter without in-phase and conjugate-phone components. Therefore, the time-domain signal is real and the optical channel will only affect its amplitude while the amplitude will be recovered at the receiver. This conjugate-symmetric OFDM modulation will reduce the throughput compared with the normal OFDM modulation with the same transceiver structure. We can adjust the spreading length to improve the physical-layer throughput. On the other hand, this conjugate-symmetric OFDM modulation carries the same symbols in both sides of the baseband, therefore the receiver signal recovery SNR can be improved by combining the symbols and its conjugate parts in the frequency domain. Since this modulation is performed at the optical carrier frequency, the Doppler computation is still based on the optical carrier frequency.

\textbf{Article Organization:}
In Sect.~\ref{sec:problem}, the research problem of configuration learning is formulated. In Sect.~\ref{sec:optical_system}, the design of the underwater optical system transceiver and its key components are provided in detail. The physical-layer frame error rate and throughput results of the system are discussed in this part. In Sect.~\ref{sec:learning}, the results of configuration learning based on the generated datasets of the underwater optical communication system are presented in this part. The conclusions are drawn in Sect.~\ref{sec:conc}.

\section{Configuration Learning Problem Formulation}\label{sec:problem}

\begin{figure*}[t]
\begin{center}
\includegraphics[width=0.8\textwidth]{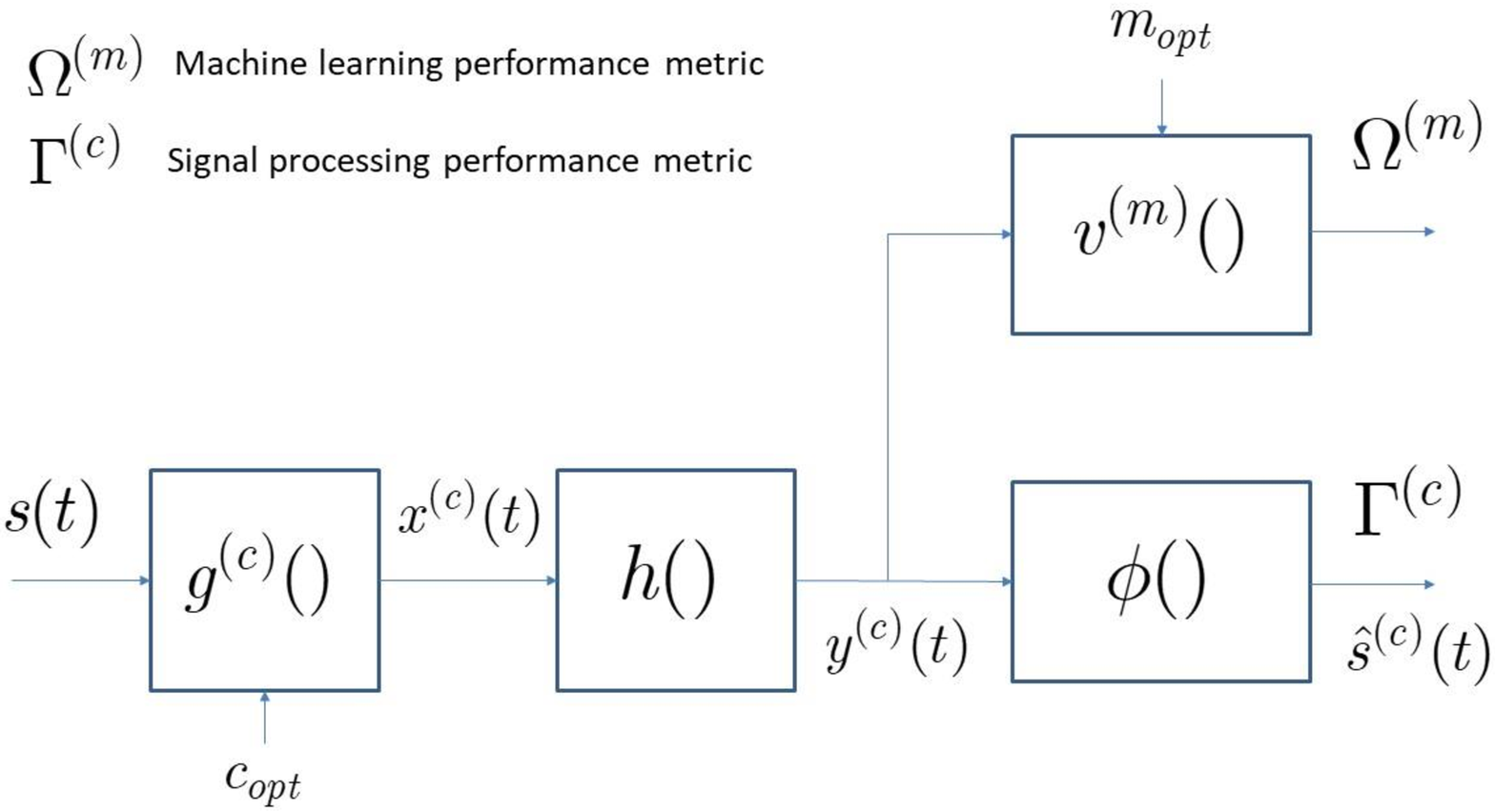}
\end{center}\vspace{-0.5cm}
\caption{Concept diagram of the configuration learning problem. The received signal is passed to two systems, one of machine-learning classifier $v^{(\bf{m})}()$ and another one of signal processing chain $\phi()$.} \label{fig:Configuration_Learning}
\vspace{-0.7cm}
\end{figure*}

{
We propose the research problem of configuration learning for general signal processing and communication systems. The proposed configuration learning research problem is depicted in Fig.~\ref{fig:Configuration_Learning}. At the transmitter/source, the signal $s(t)$ is passed to a general waveform generator $g^{(c)}()$ which can be configured to $N_C$ configurations, producing signal $x^{(c)}(t)$ with $N_C$ configurations. The $x^{(c)}(t)$ is passed to a general channel $h()$ which can be linear or non-linear, and the received signal is denoted by $y^{(c)}(t)$. At the receiver/detector, the received signal is further sent to the receiving signal processing block $\phi()$, generating a detected/decoded signal of $\hat s^{(c)}(t)$. This output signal $\hat s^{(c)}(t)$ will have a signal processing performance metric if the related performance is calculated together with the transmitted signal $s(t)$. This signal processing performance metric is denoted by $\Gamma$. The optimal configuration that maximizes the signal processing performance metric $\Gamma$ is denoted by $c_{opt}$. 


At the receiver, the received signal $y^{(c)}(t)$ is sent to a ML classifier denoted by $v^{(\bf{m})}()$, where $\bf{m}$ is a vector of the tunable parameters in the ML classifier and each tunable parameter is denoted by $m_i$. The ML classifier is to be trained by the received signal waveform $y^{(c)}(t)$, and the known optimal configuration $c_{opt}$ that maximizes the signal processing performance metric $\Gamma^{(c)}$. Then the ML classifier is deployed to classify the configuration, i.e. to find the optimal configuration $c_{opt}$ with the received signal waveform $y^{(c)}(t)$. The ML classifier $v^{(\bf{m})}()$ will produce a machine learning performance metric, the classification accuracy performance in general, denoted by $\Omega^{(\bf{m})}$ by the ML classifier of parameters $\bf{m}$. The configuration learning research problem has the following optimization objective,
\begin{equation}
    \bf{m}_{opt}  = \mathop {\max }\limits_{\bf{m}} \{ \Omega ^{(\bf{m})} \} 
\end{equation}

The concept of configuration learning outlined in this article has several critical assumptions: 1) The optimization objective is to maximize the ML classification performance $\Omega ^{(\bf{m})}$ to classify the optimal configuration $c_{opt}$, based on maximizing the signal processing metric $\Gamma^{(c)}$; 2) The variables of the optimization problem is the tunable parameters in the ML classifier $v^{(\bf{m})}()$; 3) There is no information assumed to be extracted from the receiver signal processing chain $\phi()$ sent to the ML classifier $v^{(\bf{m})}()$. 

Due to these unique assumptions, the configuration learning problem proposed in this work is a new research problem different from the research problems of rate adaptation and signal identification. These research problems have distinct optimization objectives, variables, and assumptions that are very different from our proposed configuration learning problem. Here we further elaborate on the differences: 1) In the rate adaptation research problem, the optimization objective is to maximize the physical-layer performance metric, for example, the physical-layer throughput, and the optimization variable is the configurations. In our system notation, the rate adaptation problem can be formulated by $ c_{opt}  = \mathop {\max }\limits_c \{ \Gamma ^{(c)} \}$, where $c = 1,2,...,N_C$. 
The optimization objective is the signal processing performance metric~$\Gamma ^{(c)}$, and the optimization variable is the configuration $c$ at the transmitter. Therefore, the rate adaptation problem is very different from our proposed configuration learning problem. 2) In signal identification, the optimization objective is to identify the types of the source given the received signal waveform. Given our system set up, if we assume the signal to be identified is the $x^{(c)} (t)$ with the signal type defined by $c$, this signal identification problem can also be formulated by the optimization similar to the rate adaptation formulation. The concept of signal identification is also distinguished from our configuration learning problem in which the optimization is performed on the ML performance metric~$\Omega ^{(\bf{m})}$. 


To solve the configuration learning problem requires the evaluation of the ML classifiers to obtain the ML performance metric of the classification accuracy $\Omega^{(\bf{m})}$. The theoretical investigation of ML classifier performance is the frontier of machine learning research nowadays. However, due to the non-linear nature of most ML classifiers with outstanding performance, there are no theoretical formulations developed by the research community up to date for the classifiers including LSTM-based RNN, Bi-LSTM, GRU, decision tree and AdaBoost, due to the high level of non-linearity of the structures in these ML classifiers. The research in machine learning performance evaluation is very different to signal processing/communication research, in which the error rate performance and the capacity can be preciously formulated. The most widely investigated ML theoretical performance metric, the generalization bound, is still not the performance metric to evaluate the classification accuracy of an ML classifier. The theoretical performance derivations of non-linear ML classifiers are open research problems however are not yet solved due to the high level of investigation difficulty.

Due to this research difficulty, we choose to investigate the configuration learning problem with empirical approaches. A specific signal processing/communication system, an underwater optical communication system, is investigated in this work. The waveform generator $g^{(c)}()$ is the transmitter of the underwater optical link, and the receiving signal processing block $\phi()$ is the coded receiver with Turbo code. The ML classifier $v^{(\bf{m})}()$ will be evaluated for a number of ML classifier candidates. In the following content, we will further provide the design details of this underwater optical communication system, the physical-layer performance evaluations to characterize the signal processing performance metric $\Gamma^{(c)}$ and the related $c_{opt}$ values that maximize this signal processing performance metric. The ML dataset generation method to train the ML classifier $v^{(\bf{m})}()$ is further explained with the results of the ML classification accuracy performance metric $\Omega ^{(\bf{m})}$, and the optimization of the selected ML classifiers to find optimal $\bf{m}_{opt}$ values to maximize the ML performance metric.


\begin{algorithm}[t]
\SetAlgoNoLine
\KwIn{$N_C$, $N_{dis}$, $N_{spd}$}
\KwOut{$N_{p,opt}$, $N_{h,opt}$, $u_{opt}$}
{Generate the received ML sample size $N_s = a N_{dis} N_{spd}$, where $a$ is an integer constant} \\
\For{$n$ = 1,2.,,,,$N_s$}{
 Find the optimal configuration labels $c_{opt,n}  = \mathop {\max }\limits_c \{ \Gamma^{(c)}\}$}
\For{$u$ = 1,2.,,,,$i_{MLC}$}{
$N_{p,opt} (0) = \beta$; $i = 1$; \\
\While{$|\Omega ^{(m_1, m_2)}(i+1,u) - \Omega ^{(m_1, m_2)}(i,u)|<\epsilon$}
    {Fix the number of epochs $N_{p}$, optimize the number of hidden units $N_{h}$: \\
    $N_{h,opt} (i,u) = \mathop {\max }\limits_{m_1 } \{ \Omega ^{(m_1 ,m_2 )} ,m_2  = N_{p,opt} (i - 1,,u)\}$;   \\ 
    Fix the number of hidden units $N_{h}$, optimize the number of epochs $N_{p}$: \\
    $N_{p,opt} (i,u) = \mathop {\max }\limits_{m_2 } \{ \Omega ^{(m_1 ,m_2 )} ,m_1  = N_{h,opt} (i,u)\}$;\\
    $i = i + 1$;}
} 
Determine the switched ML classifier $u_{opt} = \mathop {\max }\limits_{u} \Omega ^{(m_1, m_2)}(i,u)$ \\
Output the optimized parameters $N_{p,opt}$ = $N_{p,opt} (i,u_{opt})$; 
$N_{h,opt}$ =  $N_{h,opt} (i,u_{opt})$.
\caption{Proposed algorithm, \emph{SwitchOpt~RNN}, to address the configuration learning problem. In this design, the switching is performed among three candidate ML classifiers including LSTM, Bi-LSTM, and GRU. The tunable parameters including the numbers of epochs $N_{p}$ and hidden units $N_{h}$ are optimized by the optimization algorithm via the vector {\bf{m}} with two tunable values $m_1$ and $m_2$.}\label{alg:proposal_config_learning}
\end{algorithm}


The proposed algorithm \emph{SwitchOpt RNN} is illustrated in Algorithm~\ref{alg:proposal_config_learning}. In this algorithm, the datasets of sample size $N_s$ with each dataset containing the received signal waveform $y_n(t)$ and the optimized configuration $c_{opt,n}$ maxmizing the physical-layer throughput performance $\Gamma_n ^{(c)}$. The alternating optimization algorithm targets to optimize the number of hidden units $N_{h}$ and the associated the number of epochs $N_{p}$, by alternating the optimization of one parameter while fix another parameter. This is done by firstly intialize the number of hidden units $N_{h,opt}$ with an initial value $\beta$, then optimize the parameter $N_{p,opt}$ by maximizing the ML metric $\Omega ^{(m_1 ,m_2)}$. The optimized parameter $N_{p,opt}$ is subsequently fixed and the optimization of $N_{h,opt}$ is performed. The switching of the ML classifier is done by maximizing the ML metric among the optimized candiate ML classifiers.
}





\begin{table}[]
\small
\centering
\caption{Parameter definitions in the underwater optical system under study.}
\label{table_parameters}
\begin{tabular}{m{2.6cm}|m{10cm}}
\hline
$N_{C}$      & Number of configurations at the transmitter
\\ \hline
$c_{opt}$      & Index of the optimal configuration maximizing the performance metric
\\ \hline
$\Gamma^{(c)}$  & Performance metric of the physical-layer throughput
\\ \hline
$\Omega^{(\bf{m})}$  & Machine learning performance metric
\\ \hline
$i_{MLC}$  & Index of the ML classifier
\\ \hline
$N_{h}$ & The parameter of the number of hidden units
\\ \hline
$N_{p}$ & The parameter of the number of epochs
\\ \hline
$N_{dis}$   & Number of AUV-Buoy distances   
\\ \hline
$N_{spd}$  & Number of AUV speeds 
\\ \hline
$N_s$    &  Total number of ML samples 
\\ \hline
$\beta$ &  Initial value of the parameter $N_{p}$
\\ \hline
$m_1, m_2$ &  Two variables in the vector $\bf{m}$
\\ \hline
$N_{p,opt}$ &  The optimized number of epochs of the switched ML classifier
\\ \hline
$N_{h,opt}$ &  The optimized number of hidden units of the switched ML classifier
\\ \hline
$u_{opt}$ & The index of the switched ML classifier
\\ \hline
$\Phi_c $   & The $c$-th dataset vector where $c = 1,...,N_{C}$
\\ \hline
$N_F$      & Frequency-domain spreading code length
\\ \hline
$N_T$      & Time-domain spreading code length
\\ \hline
$N_{RX}$      & Number of photodetectors at the receiver  
\\ \hline
$N_{TX}$      & Number of LED sources at the transmitter 
\\ \hline
$N_{OFDM}$      & Number of OFDM symbols in one frame
\\ \hline
$N_{SUBC}$      & Number of data subcarriers in one OFDM symbol
\\ \hline
$N_{TFS}$      & Number of constellation symbols in one frame before time-frequency spreading; $N_{TFS}  = N_{SUBC} N_{OFDM} /(N_T N_F)$
\\ \hline
$R_C$      & Coding rate of the channel coding
\\ \hline
$M$      & Number of bits per constellation symbols modulated
\\ \hline
\end{tabular}
\vspace{-0.1in}
\end{table}


\begin{figure*}[t]
\begin{center}
\includegraphics[width=0.8\textwidth]{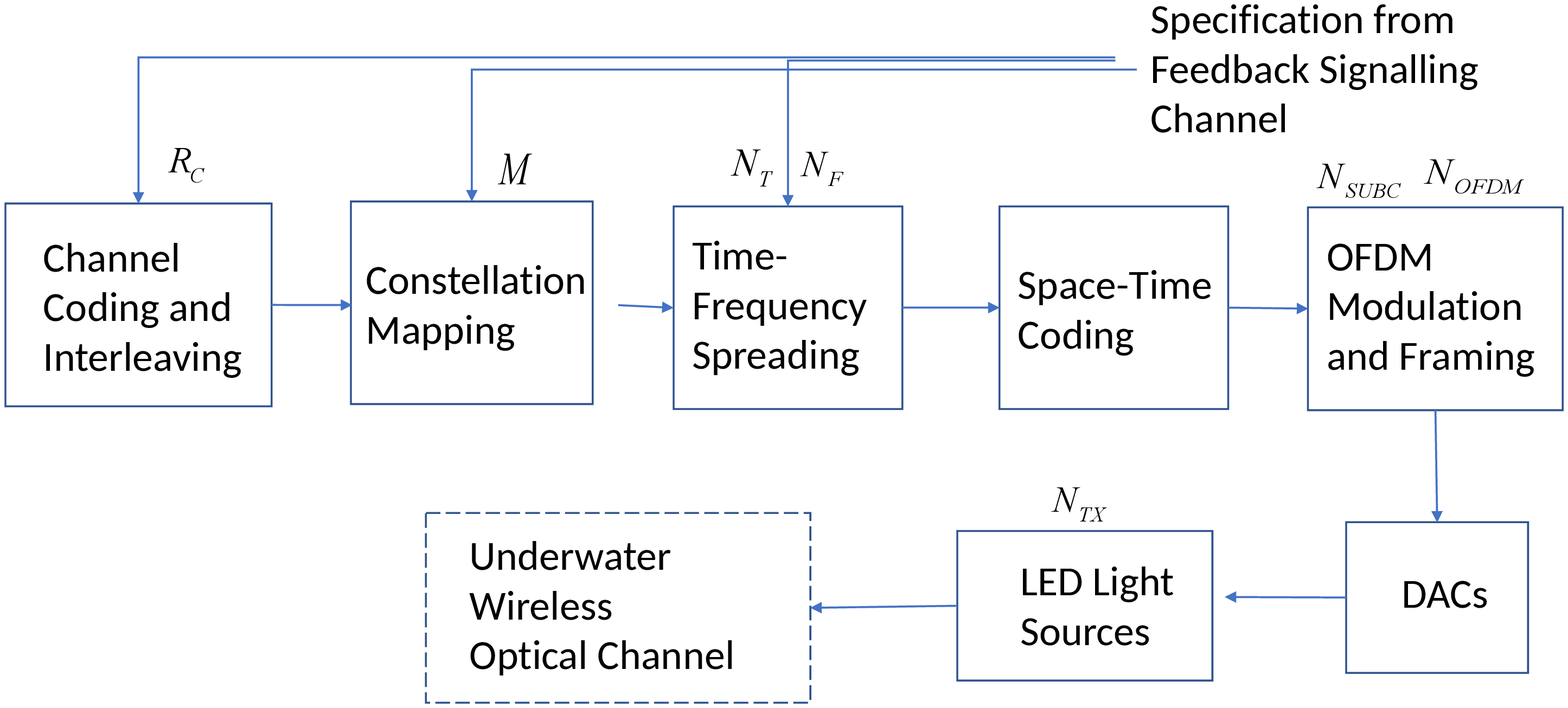}
\end{center}\vspace{-0.5cm}
\caption{{The transmitter structure of the proposed system. There is a feedback link to adjust the transmitter specification based on the ML classifier at the receiver.}}\label{fig:Tx_Structure}
\vspace{-0.7cm}
\end{figure*}

\begin{figure*}[t]
\begin{center}
\includegraphics[width=\textwidth]{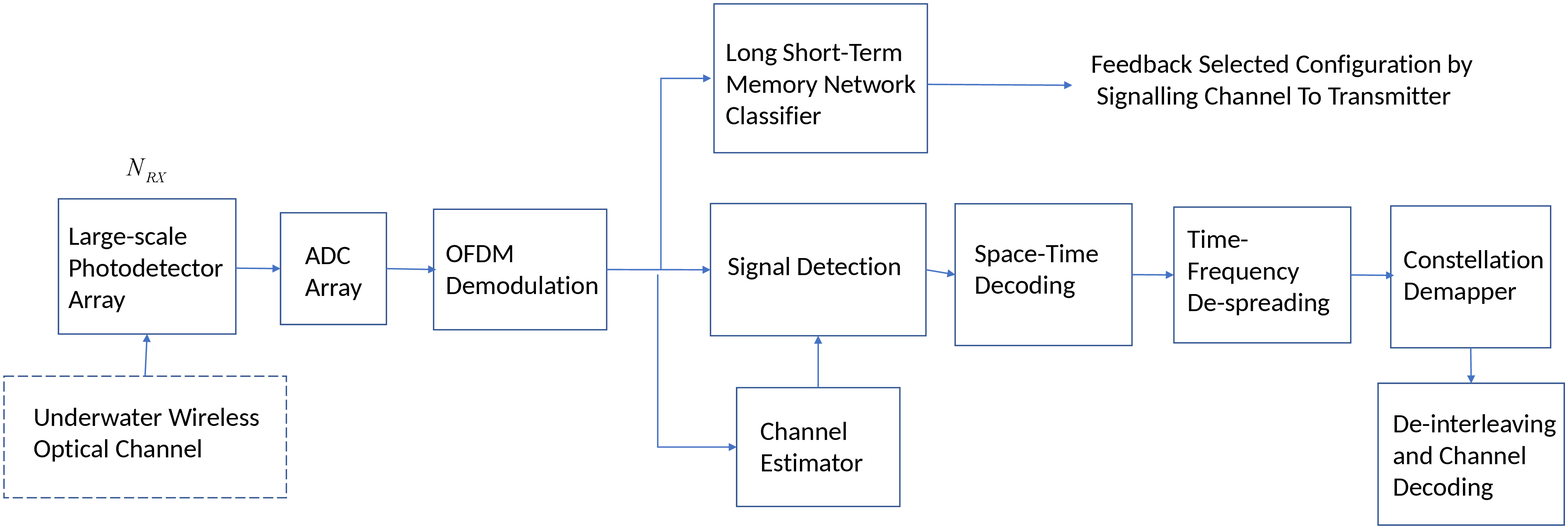}
\end{center}\vspace{-0.5cm}
\caption{Proposed configuration learning classifier at the underwater optical communication receiver. The received signal before signal detection is fed to the trained classifier for the classification of the transmitter configuration that maximizes the physical-layer throughput.} \label{fig:Rx_Structure}
\vspace{-0.7cm}
\end{figure*}

In the following content, the underwater optical communication is investigated where the configuration learning is to be designed on this particular system. The system under study is depicted in Figs.~\ref{fig:Tx_Structure} and \ref{fig:Rx_Structure} for the transmitter and receiver respectively, while the parameters are defined in Table~\ref{table_parameters}. In this system, the receiver $\phi()$ function is composed of Fast Fourier Transform~(FFT), linear filtering based detection, and Turbo code decoding. 
{The MMSE-based linear detector is adopted, and the Turbo code decoding is based on the iterative non-linear algorithm of the Logarithmic Maximum Posteriori Probability~(Log~MAP) detection~\cite{Robertson97}}.
{The ML classifier is chosen from several algorithms to maximize the accuracy of finding the optimal configuration that maximizes the physical-layer throughput.}
In the proposed system, the ML classifier is designed to identify the transmitter configuration based on the received signal waveform, as depicted in Fig.~\ref{fig:Rx_Structure}. The ML classifier is trained with the datasets of the received signal and the corresponding configuration that maximizes the physical-layer throughput during the training stage. At the online inference stage, the trained ML is deployed to learn the optimal configuration from the received data, and the optimal configuration is then fed back to the transmitter via a signaling link.

{The candidates of the ML classifier include LSTM-based RNN, Bi-LSTM, GRU, decision tree, AdaBoost, and Support Vector Machine~(SVM).} The LSTM-based RNN classifier is suited for the sequential dataset, which is the case in the signal recorded at the receiver. The LSTM neurons are non-linear in its internal structure, thus the LSTM-based RNN classifier is non-linear. {The Bi-LSTM is based on the bi-directional RNN structure with dual directions in time -- in addition to the states from the past time slots, the bi-directional RNN structure can input the state from the future time slots. This structure is a non-linear structure based on the non-linear LSTM cells with the external RNN structure of feedback of future time slots. The GRU is a non-linear RNN structure by the GRU cell, which removes the output gate and reduces the complexity compared with LSTM cell, and is shown to have better performance than LSTM in certain datasets of small size.} The decision tree classifier is a non-linear classifier and the classification process is based on the sorting and searching on a tree structure, which is a non-linear operation. The AdaBoost classifier combines multiple non-linear classifiers, thus it is also non-linear. {In contrast, the support vector machine~(SVM) is a linear classifier. There is recent work discussing the theoretical analysis of the performance of the SVM. However, for the non-linear classifiers, there are no existing theoretical formulations of the classification accuracy, due to the high level of investigation difficulty. Nevertheless, in our empirical investigation the classification performance of the non-linear classifiers of LSTM, Bi-LSTM and GRU outperform other competing ML classifiers for the proposed configuration learning problem. Our work is the first proposal to optimize the non-linear ML classifiers of LSTM, Bi-LSTM and GRU by the alternating optimization approach based on the physical-layer configuration optimization, thus this research problem is named the \it{Configuration Learning} problem.
}

{
We input the received signal waveform for the training of the configuration learning classifier; the amplitude of the signal at the receiver is extracted and applied to the configuration learning classifier training. The datasets are generated by a bit-level simulator that emulates the physical-layer performance in realistic oceanic environments. The datasets are generated by the following procedure:
\begin{enumerate}
\item The transmitter configurations are defined to be the combinations of channel coding rate $R_C$, length of time-frequency spreading $N_T$ and $N_F$. Assume the transmitter has $N_{C}$ configurations.  
    
\item The distances between transmitter and receiver are varied for $N_{dis}$ points; For each distance point, the AUV transmitter speed is varied for $N_{spd}$ points. The physical-layer throughput of $N_{C}$ configurations are simulated for these $N_{C} = N_{dis} N_{spd}$ points. The obtained results are the physical-layer throughput results. For the $c$-th point, $c = 1,...,N_{C}$, the index of the configuration that maximizes the physical-layer throughput is $c_{opt}$, selected from $N_{C}$ number of configurations.
    
\item The received signal waveform is recorded in $N_{RX}$ receiving elements, for the $N_{dis} N_{spd}$ points. The $c$-th dataset is denoted by $\Phi _c $, ($c = 1,...,N_{C}$). For each dataset $\Phi _c $, there is an optimal index of the configuration $c_{opt}$. The pairs of $\Phi _c $ and optimal index $c_{opt}$ are recorded as one ML sample.
\end{enumerate}

{In the generated datasets, $N_{dis} = 240$ and $N_{spd} = 4$. There are $960$ ML samples generated with four AUV speeds of $0.1$, $0.3$, $0.4$, and $0.5~\rm{m/s}$. For each AUV speed, there are $240$ ML samples, each distance value with $4$ samples. The evaluated distances range from $1$ to $60~\rm{m}$. For each distance, 1-by-4 MIMO setup is simulated with $32$ complex numbers in one OFDM symbol, resulting in a real-number vector $\Phi _c $ of length $128$.} 
The labels of the ML samples are the corresponding spreading and coding schemes $c_{opt}$ that maximize the physical-layer throughput. The dataset generation is depicted in Table~\ref{table_data} with the explanation of the dataset generation in one table.

\begin{table*}[t]
\centering
\caption[loftitle]{Dataset generated by the physical-layer simulator to evaluate the ML classifiers.}\label{table_data}
\centering
\begin{tabular}{m{5cm}|m{11cm}}
\hline
 Variable AUV speed    & $0.1$, $0.3$, $0.4$, and $0.5~\rm{m/s}$  \\ \hline
 Variable distance     & 60 AUV-Buoy distances, from $1$ to $60~\rm{m}$, with spacing of $1~\rm{m}$ \\
for each AUV speed & Totally $960$ ML samples \\ \hline
ML feature vector  & Vector of $128$ complex numbers \\
for each ML sample   & $32$ per OFDM symbol multiplied by $4$ receiving photodetector \\ \hline
  ML label  & The coding and spreading configuration maximizing the \\
  for each ML sample & physical-layer throughput \\ \hline
 Generated dataset size & $960$ samples of feature vector \\
  & $960$ samples of label \\ \hline
\end{tabular}

\end{table*}

}

\section{Proposed Optical System Design and Performance Results}\label{sec:optical_system}






The transmitter of the proposed system is shown in Fig.~\ref{fig:Tx_Structure}. In our system, we apply Turbo code as channel coding with a code rate of $1/2$ or $1/3$, after which the interleaving technique is adopted to make the error bits in short deep fading sparsely placed in the data sequence. On modulation, we adopt QPSK to improve the bit rate. Then, we do the serial-to-parallel conversion and apply the time-frequency domain spreading with Hadamard code. {After we map the symbols to transmit antennas with space-time coding, we generate OFDM frames in a conjugate-symmetric way.} The receiver simulator structure is shown in Fig.~\ref{fig:Rx_Structure}. Here we apply a large scale of receiving photodetectors. The channel estimation outputs at the pilot OFDM symbols are applied for the Maximum Ratio Combining~(MRC), which improves the receiver sensitivity for the system of single or multiple optical transmitters and massive MIMO photodetector, with Zero-Forcing~(ZF) detection at the receiver for signal recovery.

To improve the communication coverage when using optical waves underwater, we propose a new approach that enhances the sensitivity of the receiver by deploying a large-scale photodetector array. Considering the low cost of photodetectors, it is feasible to design an array of hundreds of photodetectors at the receiver. Moreover, to mitigate the Doppler effect caused by the mobile AUV and to improve the SNR at the receiver, time-frequency spreading is designed using orthogonal codes (e.g., Hadamard code). Note that OFDM modulation is utilized to improve spectrum efficiency. The system is implemented by a physical-layer simulator with underwater optical communication channel. The frame structure is composed of one pilot OFDM symbol for channel estimation, followed by one data OFDM symbol for data transmission. In the pilot symbols, a pilot placement scheme named interlaced pilot~\cite{Zhao06} is adopted as the transmitting pilot design for the MIMO scheme. In this pilot design, different transmitter places the pilot in a spatially orthogonal way such that, at one subcarrier, there is only one pilot from one transmitter. This design can greatly simplify the channel estimator algorithm at the receiver while keeping low computational complexity.

\subsection{Underwater Wireless Optical Channel Model Review}\label{sec:channel}

In this part, the channel and noise models of the underwater optical systems are reviewed as the basis of the evaluations. 
There are two models for underwater optical communication channels: 1)~large-scale path loss model, which is described by a set of deterministic equations; 2)~small-scale fading of the multipath time-varying channel, which includes the effects of scattered lights. Regarding the large-scale path loss model, the channel path loss and four types of noises are modeled by the reported empirical modeling. The underwater optical channel model to capture the scattering and absorption effects has been proposed in~\cite{Jaruwatanadilok08}. The channel impulse response caused by the scattering is modeled via a double Gamma function for the underwater optical channel~\cite{Tang14}. 
The underwater noise models have been discussed in works~\cite{Anguita11,Jaruwatanadilok08}. The received signal strength model has been discussed in works~\cite{Doniec13,Kaushal16,Anguita11}. The underwater optical noise is composed of four parts: i)~the solar background noise, ii)~the shot noise produced by the incident received optical light, iii)~the leakage/dark current noise and iv)~the thermal noise. 

The power of the current $i_S^2$ by the solar background noise in the unit of $\rm{A}^2$ can be expressed by,
$i_S^2  = \left( {\Phi_s A_r \gamma_0 } \right)^2$,
where $\Phi_s$ is the scalar irradiance of the solar light in $\mathrm{W/m^2}$; $A_r$ is the receiving area of the photodetector in $\mathrm{m^2}$, and $\gamma_0$ is the receiver sensitivity in $\mathrm{A/W}$.
The power of the shot noise current $i_L^2$ produced by the incident received optical light is modeled as,
$i_L^2  = 2q I_L B$,
where $q$ is the electronic charge of $1.6 \times 10^{-19}~\rm{C}$, $I_L$ is the photocurrent parameter, and $B$ is the electronic bandwidth. The power of the leakage/dark current noise $i_D^2$ in the unit of $\rm{A}^2$ is,
$i_D^2  = 2q I_D B$,
where $I_D$ is a photodiode parameter. The thermal noise can be expressed by,
$i_T^2  = 4KTB/R$,
where $K$ is the Boltzmann constant of $1.38 \times 10^{-23}~\rm{J/K}$, $T$ is the temperature in Kelvin, and $R$ is the load resistance.  
For the direct Line Of Sight~(LOS) link, the received optical signal strength can be modeled as,
$i_R^2  = P_T \eta_0 L_a\frac{{A_r \cos \beta_0 }}{{2\pi d^2 (1 - \cos \theta )}}$,
where the attenuation $L_a$ is further expressed as,
$L_a = \exp \left\{ { - c_a(\lambda_0 )d} \right\}$.
In the above expressions, $P_T$ is the transmitting source-radiated optical power in $\rm{W}$; $\eta_0$ is the combined efficiency of the transmitter and receiver; $\beta_0$ is the inclination angle of the receiver w.r.t. the light beam; $\theta$ is the diverge angle of the transmitter optical beam; $d$ is the transmitter-receiver distance; $c_a$ is the attenuation coefficient in $\rm{m}^{-1}$ for the light beam at the wavelength of $\lambda_0$. 
The SNR of the underwater optical link at a distance of $d$ between the transmitter and receiver can be expressed by,
$SNR = i_R^2 / (i_S^2 + i_L^2 + i_D^2 + i_T^2)$.
The wavelength of $450-500~\rm{nm}$ is ideal for pure sea/clear water, whereas the wavelength of $520-570~\rm{nm}$ is ideal for the coastal ocean and turbid harbor water~\cite{Kaushal16}. This is because the absorption and scattering effects differ due to the varied chemical concentration at different locations of the sea.

{
Regarding the small-scale fading of the multipath time-varying channel in underwater, there are works on the measurements of the multipath delay. The $10~\rm{ns}$ multipath delay is measured in~\cite{Lacovara08} and~\cite{Lanzagorta12} for an underwater optical communication channel {within a distance up to $100~\rm{m}$}. The multipath is caused by the scattered lights in the underwater propagation environments. 
In the simulation and OFDM-based optical transceiver design, the channel assumptions include the $10~\rm{ns}$ multipath delay and time-varying Doppler effects. The Doppler effect is caused by the AUV moving and is modeled by the AUV moving speed. We have assumed multiple AUV speed including $0.1$, $0.3$, $0.4$, and $0.5~\rm{m/s}$.}

\subsection{OFDM-based Optical Transceiver Design}

{
To mitigate ISI caused by the multipath delay, the OFDM modulation is adopted in the system design. The multipath of $10~\rm{ns}$ will introduce ISI for an optical communication system of direct On-Off Keying~(OOK) modulation. For $1~\rm{GHz}$ sampling rate, the OOK symbol interval is $1~\rm{ns}$. The $10~\rm{ns}$ multipath will introduce ISI spanning about ten OOK symbols.
Due to the ISI with OOK modulation of $1~\rm{GHz}$ sampling rate for $10~\rm{ns}$ multipath, there is a strong motivation to adopt the OFDM modulation to combat the ISI and to improve the performance of the system. The OFDM parameter design includes the choice of Cyclic Prefix~(CP) length to be greater than the multipath delay of $10~\rm{ns}$ so as to avoid interference between the OFDM symbols.  

{In our proposal, the modulation is done by adjusting the optical source intensity by the time-domain OFDM symbol amplitude by conjugate-symmetric OFDM modulation.} The scattering and turbulence of the underwater channel will introduce distortion in the signal, which is part of the wireless optical channel effect. The solution provided in our design to address the wireless optical channel effect consists in proposing a complete physical-layer system design, including the frame structure, the pilot design, and the receiver signal processing chain. The pilots in the OFDM frame are able to estimate the channel response; the transmitted signal is recovered by the signal detection block, thus mitigating the scattering and turbulence effects. In brief, this complete physical-layer system design is able to solve the scattering and turbulence that will distort the coherent signal intensity modulation. Therefore, the coherent modulation scheme of QAM with OFDM can work well in the proposed system. This is one of the contributions of the proposal to the underwater optical communication system design. 
}

In existing optical systems adopting OOK modulation, very-high-speed Analog to Digital Converters~(ADC) up to $10~\rm{Gbit/s}$ are designed. However, this high-sampling-rate ADC does not apply to our system due to the resolution bits and the number of ADCs needed in our system. Because the OFDM signals are to be digitalized at the receiver, the ADC resolution bits of our design should be much higher than in an OOK system. Additionally, in our system, there are a large number of photodetectors, and each photodetector needs an ADC, therefore the total number of ADCs required is a large number. Due to the high cost of $\rm{GHz}$-sampling-rate ADC, a large number of such ADCs is not feasible be designed in the massive MIMO receiver. We can only adopt the ADC of the medium sampling rate. It can be assumed that the ADC has a sampling rate of $100~\rm{MHz}$. The length of the cyclic prefix is designed to be longer than the maximum multipath delay to avoid ISI. In our system, the frame structure is composed of OFDM symbols with partial subcarriers allocated to the pilot and the rest of the subcarriers allocated to data.

\begin{table}
\small
\caption[loftitle]{Simulation Parameter Specifications.}\label{tbl:table_parameters}
\centering
\begin{tabular}{m{10.6cm}|m{3.0cm}}\hline
\textbf{Parameter} &\textbf{Value} \\
\hline
Carrier frequency		& $475~\rm{THz}$                              \\ \hline
Bandwidth of OFDM carrier $B$        & $100~\rm{MHz}$      \\ \hline
OFDM FFT size                     & $32$   \\ \hline
OFDM symbol length         & $340~\rm{ns}$    \\ \hline
OFDM cyclic prefix length        & $20~\rm{ns}$   \\ \hline
Number of optical sources at AUV $N_{TX}$                   & $1$   \\ \hline
Massive MIMO number of photodetectors $N_{RX}$  & $100$        \\ \hline
Frequency-domain spreading length $N_F$  & $16$                  \\ \hline
Time-domain spreading length $N_T$  & $8$         \\ \hline
Turbo code rate $R_{C}$ & $1/2$ or $1/3$                               \\ \hline
AUV speed & $0.1-0.5~\rm{m/s}$                               \\ \hline
Length of one frame & $160$ OFDM symbols                               \\ \hline
Scalar irradiance  of  the  solar light $\Phi_s$ & $0.8109~\rm{W/m^2}$ \\ \hline
Receiver aperture area $A_r$                    & $0.01~\rm{m^2}$   \\ \hline
Transmitting power at an AUV $P_T$					&$50~\rm{W}$\\ \hline
Receiver sensitivity $\gamma_0$       & $0.5~\rm{A/W}$   \\ \hline
Combined optical efficiency of transmitter and receiver $\eta_0$     & $81\%$  \\ \hline
Transmitter inclination angle $\beta_0$ & $0^{\circ}$    \\ \hline
Laser beam divergence angle $\theta$					& $68^{\circ}$ \\ \hline
Extinction coefficient, clear ocean	$c_a$			&$0.1514~\rm{m^{-1}}$\\ \hline
Photocurrent  parameter $I_L$ & $100$ \\ \hline
Temperature $T$ & $290~\rm{K}$ \\ \hline
Load resistance $R$ & $100~\rm{\Omega}$ \\ \hline
Photodiode parameter $I_D$ & $1.226 \times 10^{-9}$ \\ \hline
\end{tabular}
\end{table}

The choices of OFDM parameters depend on Doppler and multipath. Given the supported speed, we can calculate the Doppler and coherent time. The FFT duration of the OFDM symbol must be much less than the coherence time of the channel so that there is no inter-subcarrier interference. Considering laser at $450$ to $550~\rm{nm}$ wavelength, given the optical propagation speed of $2.26 \times 10^{5}~\mathrm{km/s}$ in water, the carrier frequency of the laser spectrum in this above wavelength range is $451$ to $501~\rm{THz}$. We choose the carrier frequency of $475~\rm{THz}$ for the proposed system. Assuming a AUV speed of $0.1~\rm{m/s}$ and a carrier frequency of $475~\rm{THz}$, the coherence time is $4758~\rm{ns}$. Therefore the OFDM FFT size is chosen to be 32 with an FFT duration of $320~\rm{ns}$, much smaller than the coherence time, to ensure that there is no inter-subcarrier interference. Therefore, the OFDM FFT size is chosen to be $32$ with an FFT duration of $320~\rm{ns}$; the Cyclic Prefix~(CP) length is two samples of duration $20~\rm{ns}$ to mitigate the multipath effect. Thus, overall, the OFDM symbol length is $340~\rm{ns}$. {Since we apply the conjugate-symmetric OFDM modulation, there are $16$ different symbol streams in the $32$ subcarriers.}

Due to the laser wavelength of less than one micrometer, a large number of optical elements can be integrated into a phone-sized space. We assume a square area of integrated photodetectors. 
On the definition of the massive MIMO receiver, it is generally assumed that the number of elements is at least one order of magnitude greater than the {usual} MIMO case so that the ergodic capacity of massive MIMO can be much higher than in the MIMO scenario. There is work studying for radio communication on how to select the number of elements for massive MIMO setup~\cite{Hoydis13}. For massive MIMO in this underwater optical communication application, we assume 100 (10-by-10) photodetectors at the receiver. The selected major optical OFDM and channel parameters are summarized in Table~\ref{tbl:table_parameters}. It is further assumed that the spacing between the photodetectors is much larger than the wavelength so that the channel realizations can be assumed independent among the receiving photodetectors.

\balance

 \begin{figure*}[t]
  \begin{minipage}[ht]{0.33\linewidth}
 \centering
 \includegraphics[width=\textwidth]{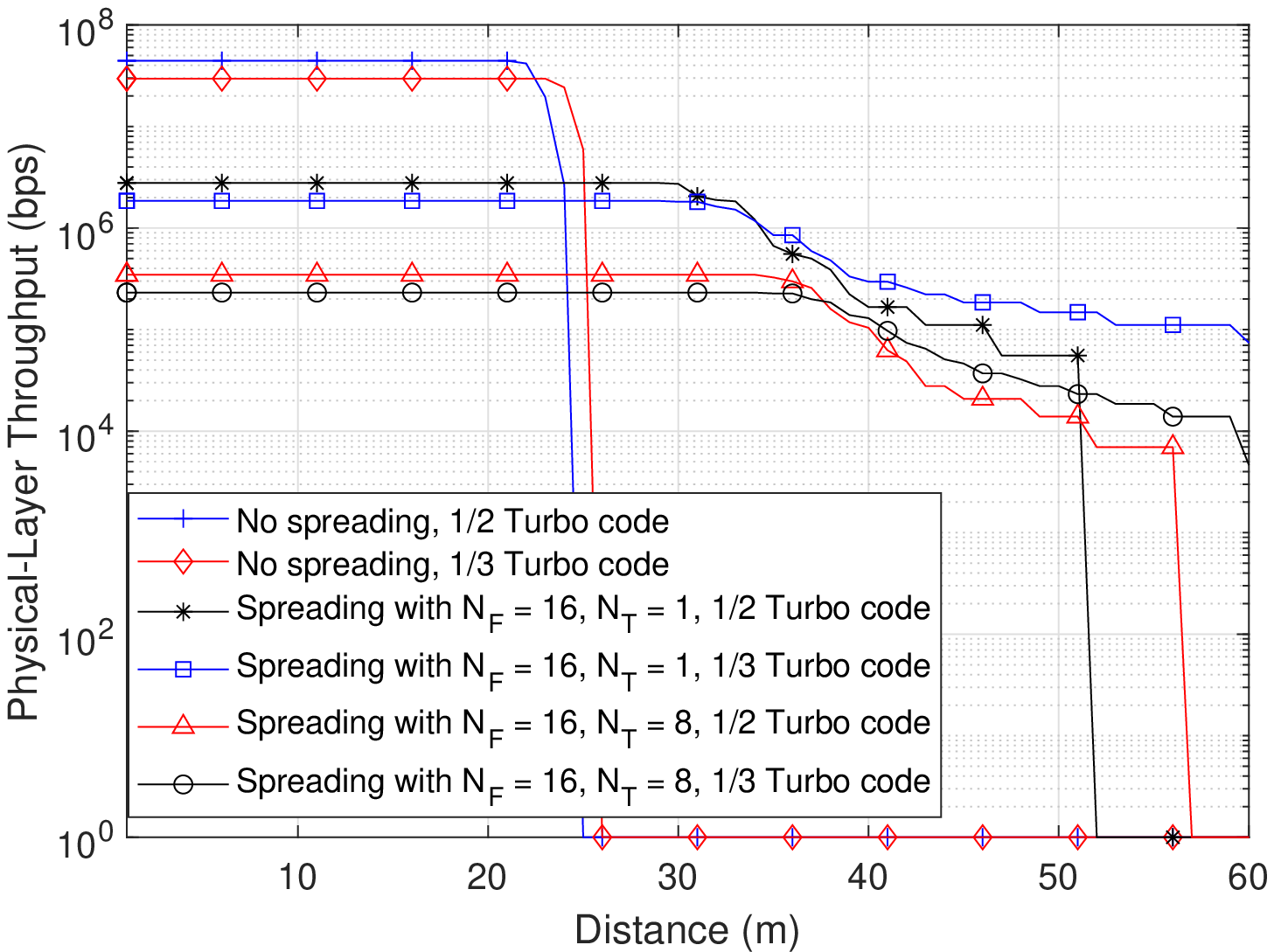}
 \begin{center} 
 (a)
 \end{center}
 \end{minipage}%
 \begin{minipage}[ht]{0.33\linewidth} 
 \centering
 \includegraphics[width=\textwidth]{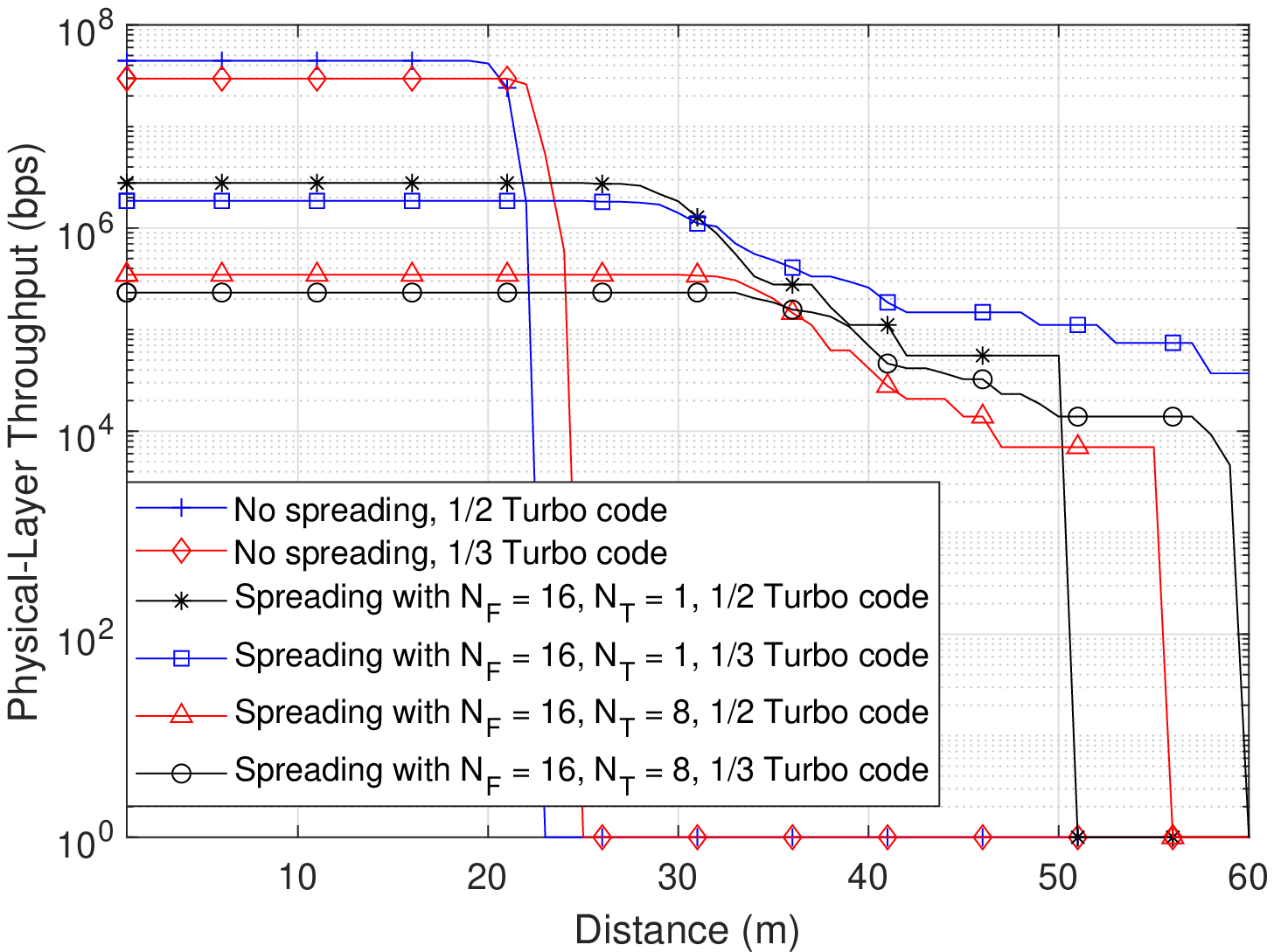}
 \begin{center}
 (b) 
 \end{center} 
 \end{minipage}
 \begin{minipage}[ht]{0.33\linewidth} 
 \centering
 \includegraphics[width=\textwidth]{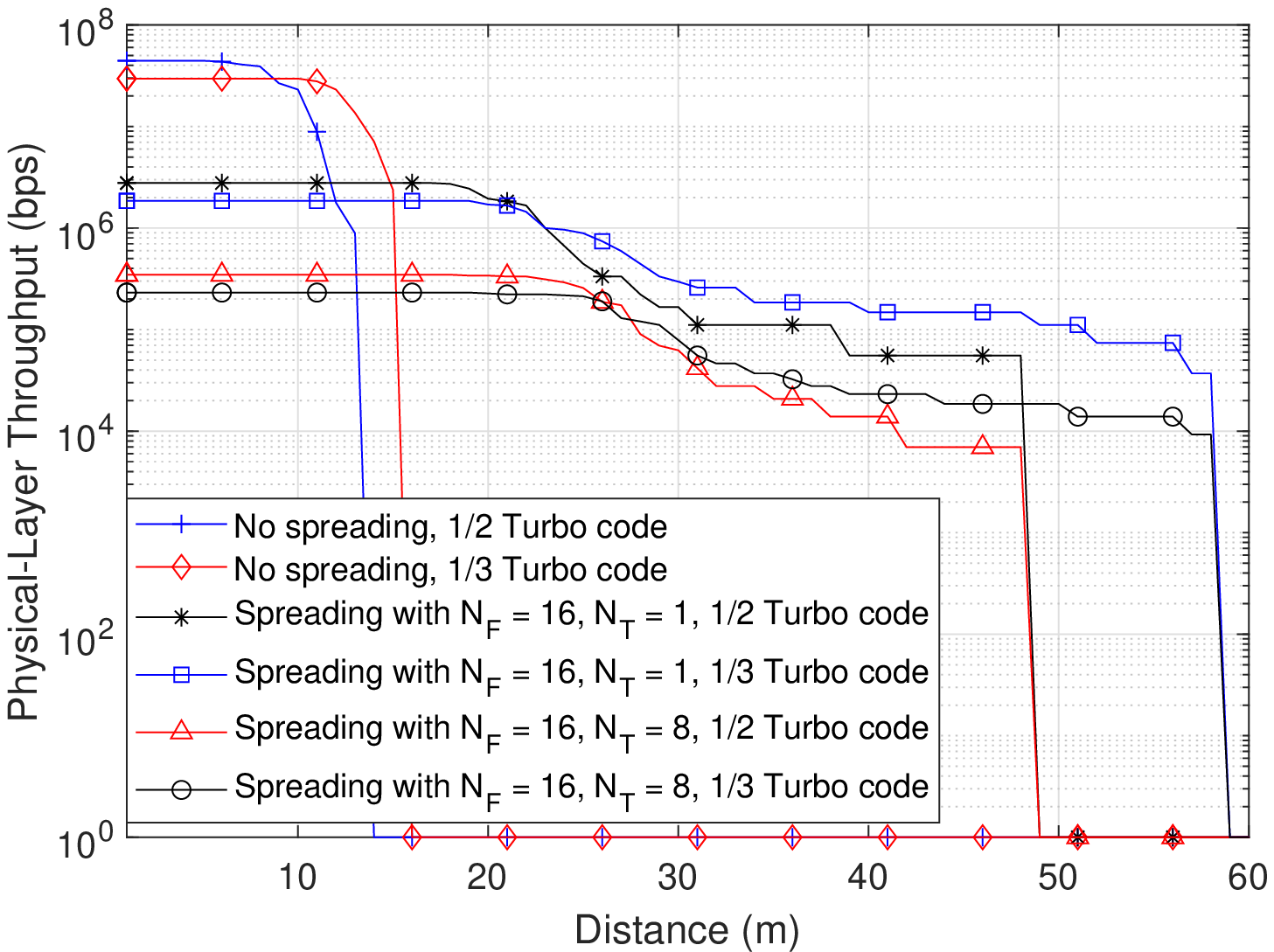}
 \begin{center}
 (d) 
 \end{center} 
 \end{minipage}
\caption{Physical-layer throughput for 1-by-100 QPSK with an AUV speed of (a) $0.1~\rm{m/s}$, (b) 
$0.3~\rm{m/s}$, (c) $0.5~\rm{m/s}$.
}\label{fig SIMO TF PHY}
 \end{figure*}

\begin{figure*}[t]
  \begin{minipage}[ht]{0.33\linewidth}
 \centering
 \includegraphics[width=\textwidth]{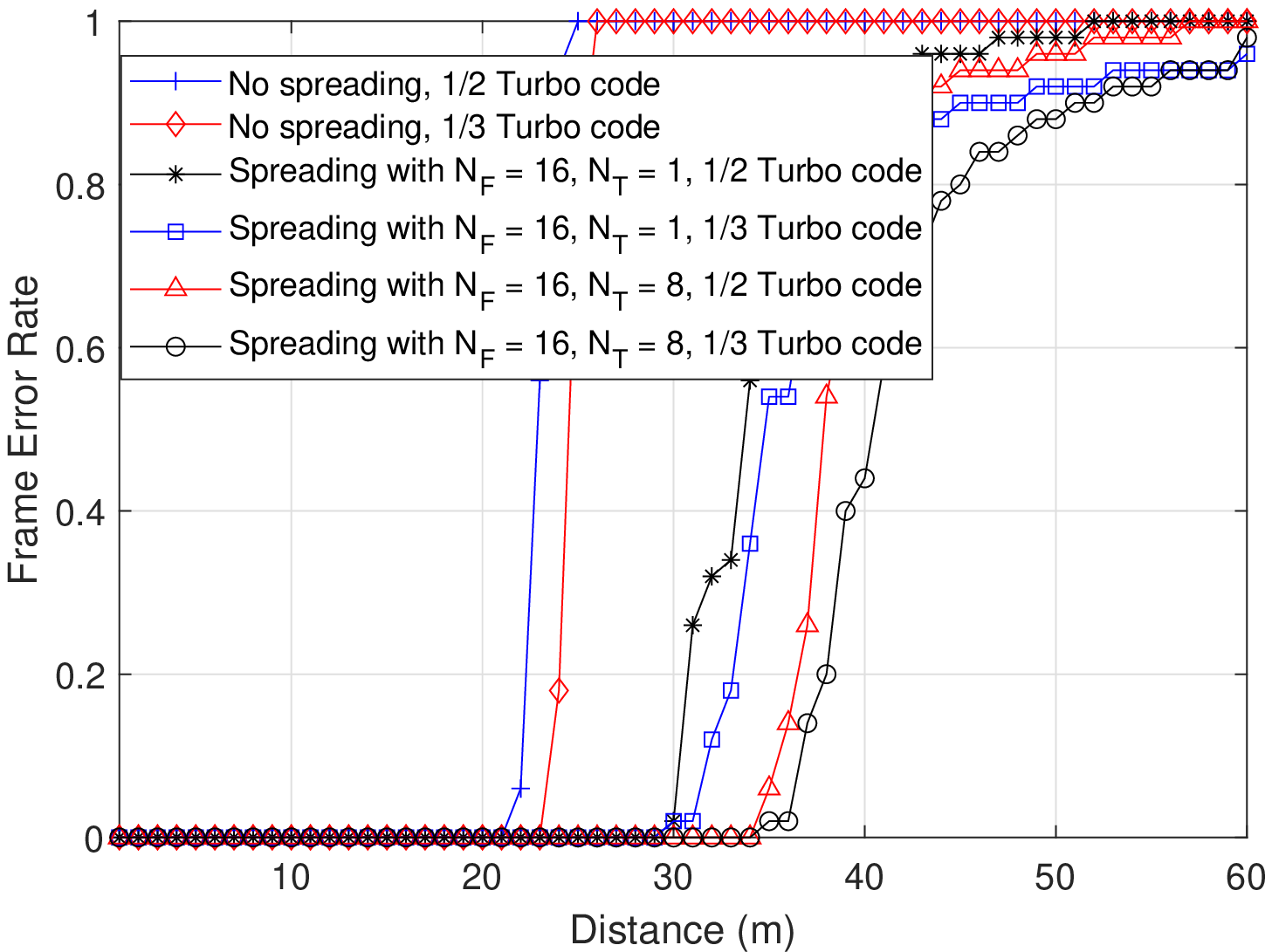}
 \begin{center} 
 (a)
 \end{center}
 \end{minipage}%
 \begin{minipage}[ht]{0.33\linewidth} 
 \centering
 \includegraphics[width=\textwidth]{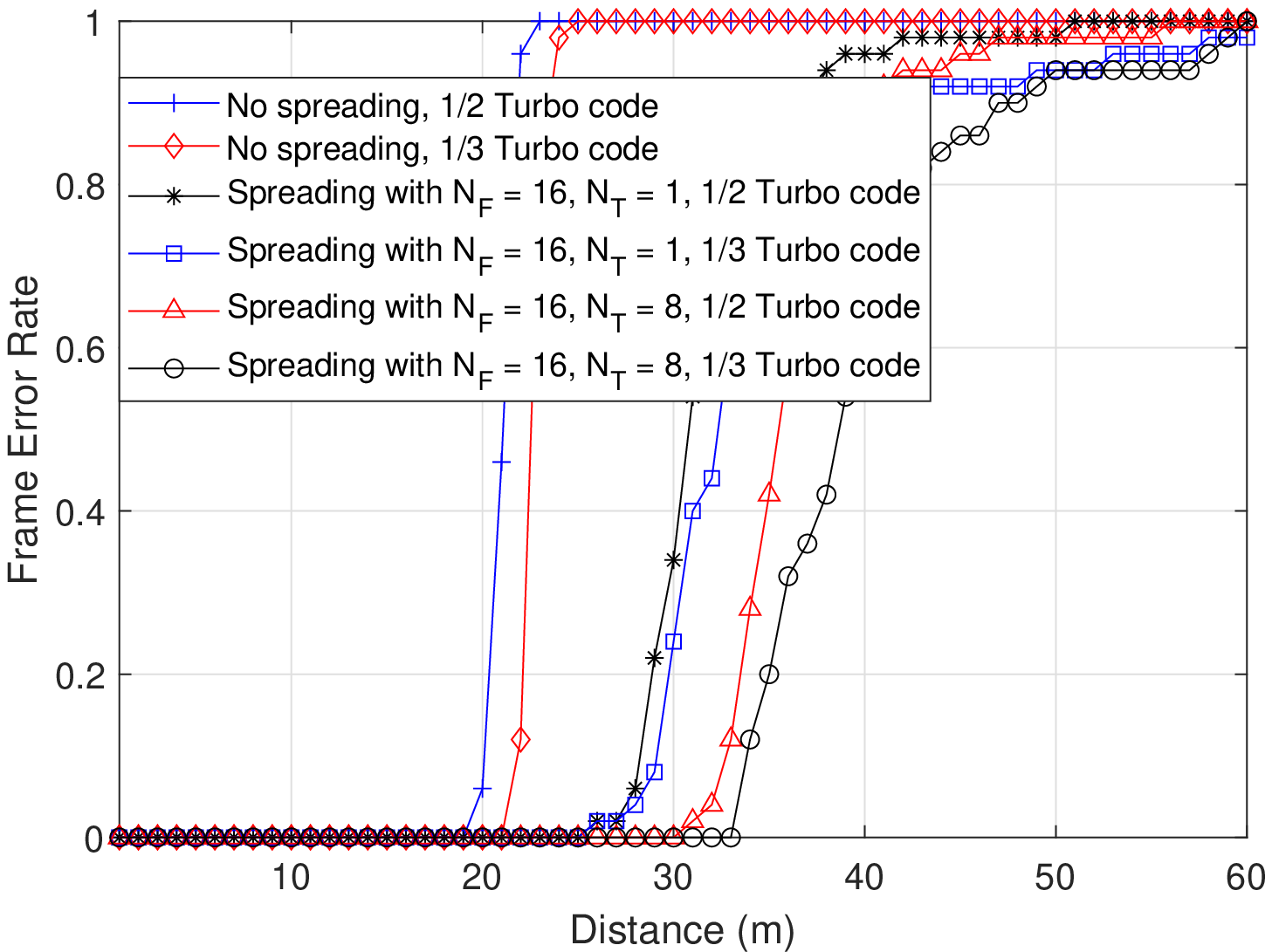}
 \begin{center}
 (b) 
 \end{center} 
 \end{minipage}
 \begin{minipage}[ht]{0.33\linewidth} 
 \centering
 \includegraphics[width=\textwidth]{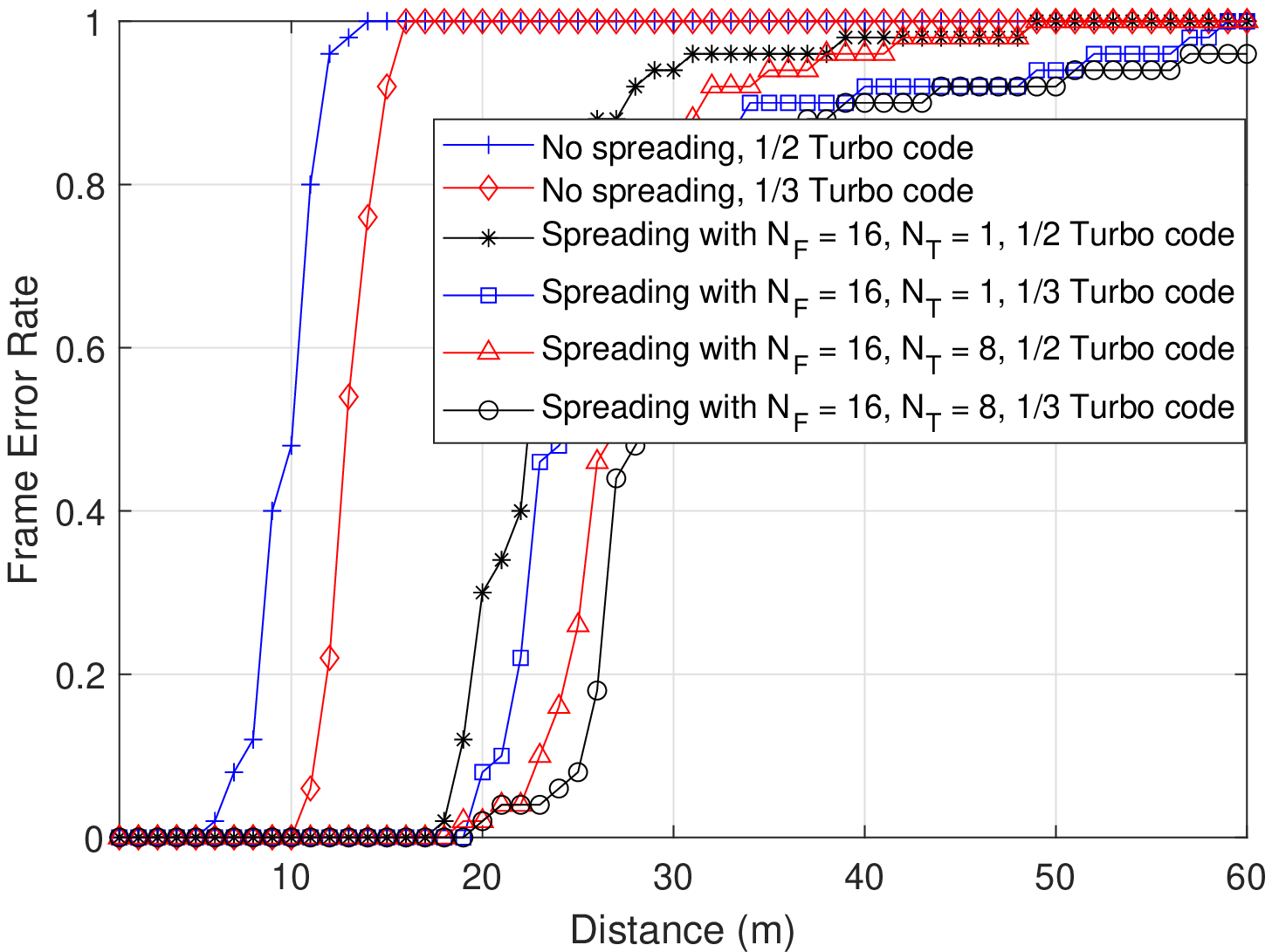}
 \begin{center}
 (c) 
 \end{center} 
 \end{minipage}
\caption{Physical-layer frame error rate for 1-by-100 QPSK with an AUV speed of (a) $0.1~\rm{m/s}$, (b) 
$0.3~\rm{m/s}$, (c) $0.5~\rm{m/s}$.
}\label{fig SIMO TF FER}
 \end{figure*}
 
\begin{figure*}[t]
  \begin{minipage}[ht]{0.33\linewidth}
 \centering
 \includegraphics[width=\textwidth]{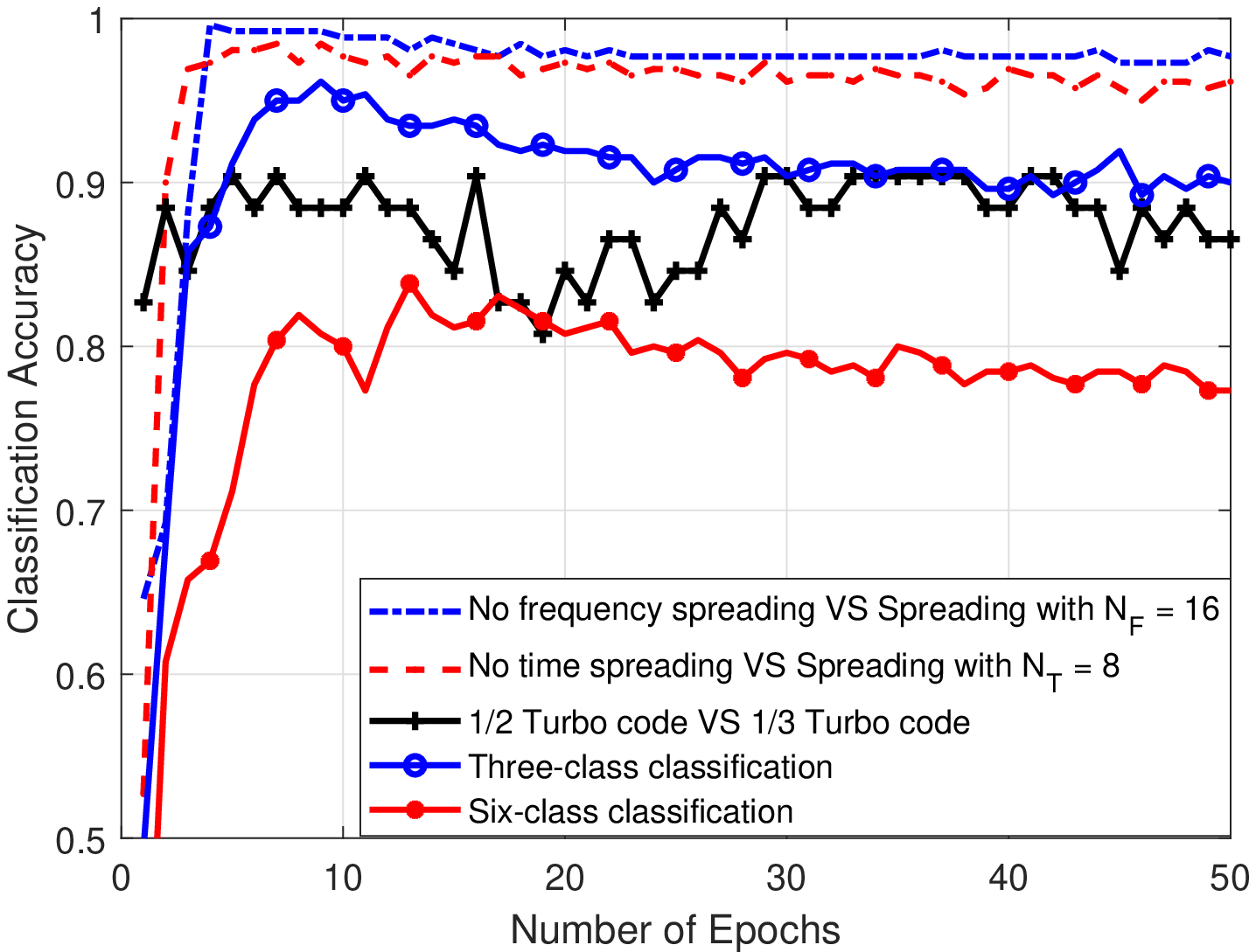}
 \begin{center} 
 (a)
 \end{center}
 \end{minipage}%
 \begin{minipage}[ht]{0.33\linewidth} 
 \centering
 \includegraphics[width=\textwidth]{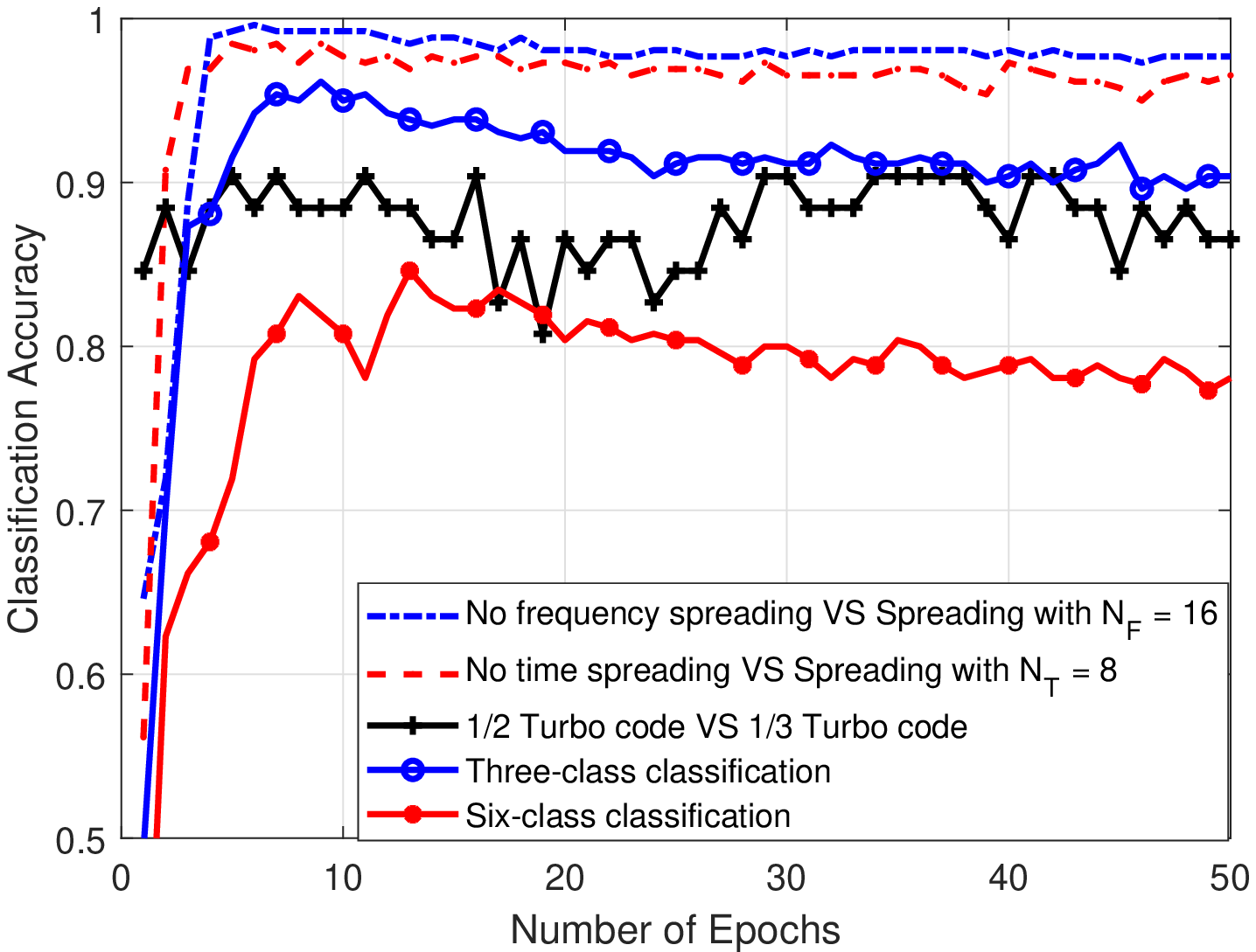}
 \begin{center}
 (b) 
 \end{center} 
 \end{minipage}
 \begin{minipage}[ht]{0.33\linewidth} 
 \centering
 \includegraphics[width=\textwidth]{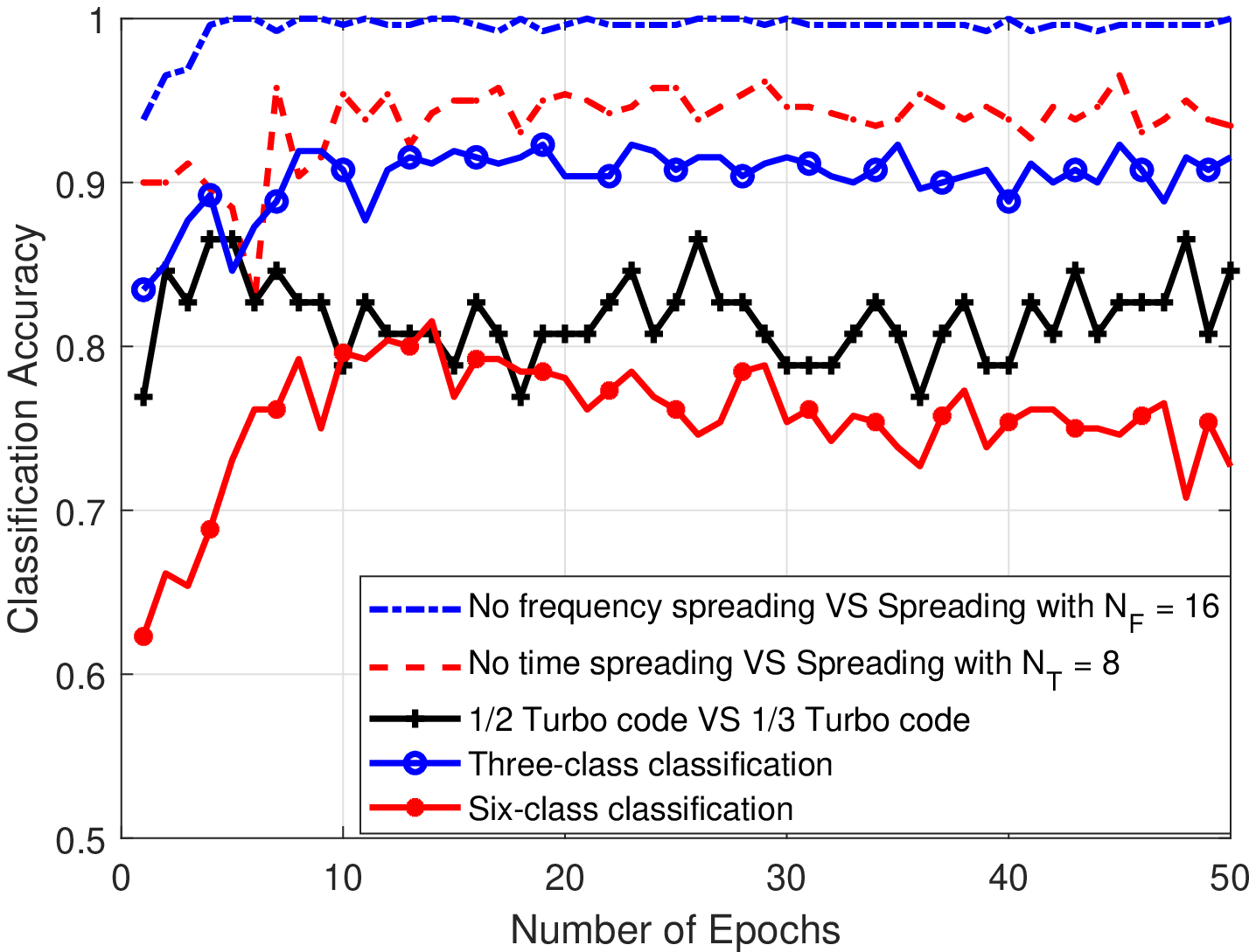}
 \begin{center}
 (c) 
 \end{center} 
 \end{minipage}
 
  \begin{minipage}[ht]{0.5\linewidth} 
 \centering
 \includegraphics[width=0.66\textwidth]{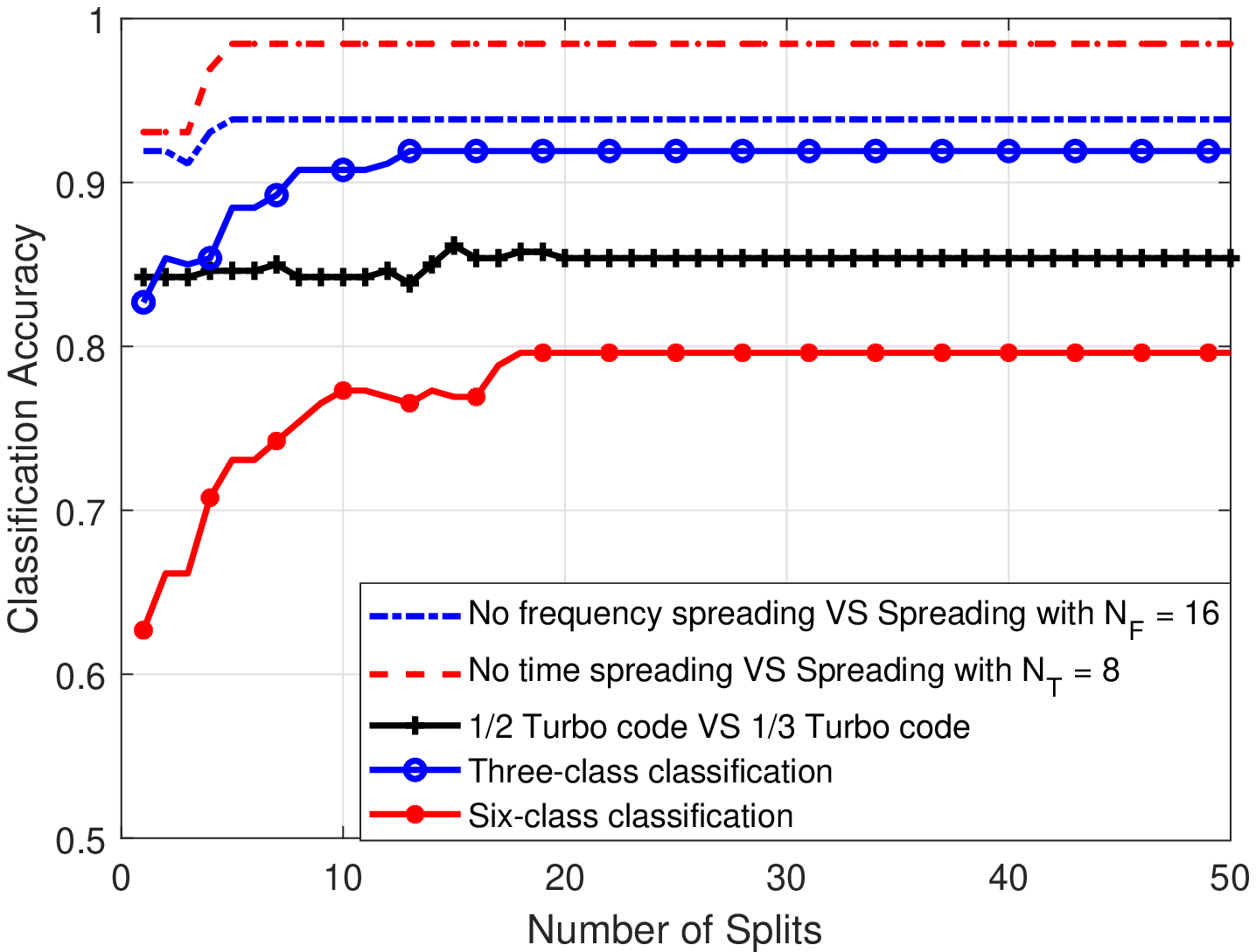}
 \begin{center}
 (d) 
 \end{center} 
 \end{minipage}
 \begin{minipage}[ht]{0.5\linewidth} 
 \centering
 \includegraphics[width=0.66\textwidth]{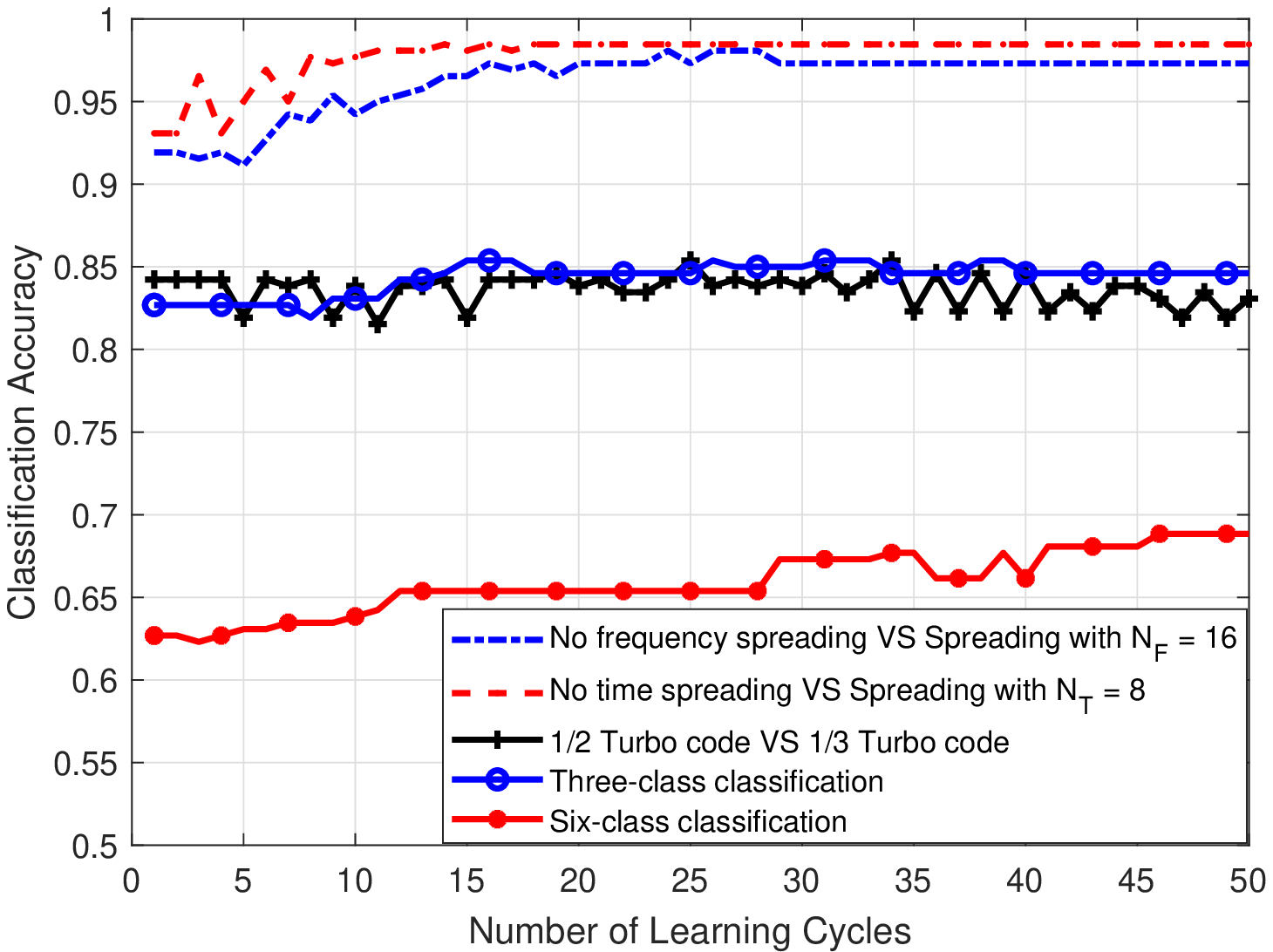}
 \begin{center}
 (e) 
 \end{center} 
 \end{minipage}
\caption{Training classification error convergence curve of classifiers: (a) LSTM-based RNN classifier with $600$ hidden units; (b) Bi-LSTM classifier with $600$ hidden units; (c)  GRU classifier with $600$ hidden units; (d) decision  tree  classifier; (e) adaptive  boosting  ensemble  classifier.
}\label{fig ML classifier}
 \end{figure*}
\begin{figure*}[t]
  \begin{minipage}[ht]{0.33\linewidth}
 \centering
 \includegraphics[width=\textwidth]{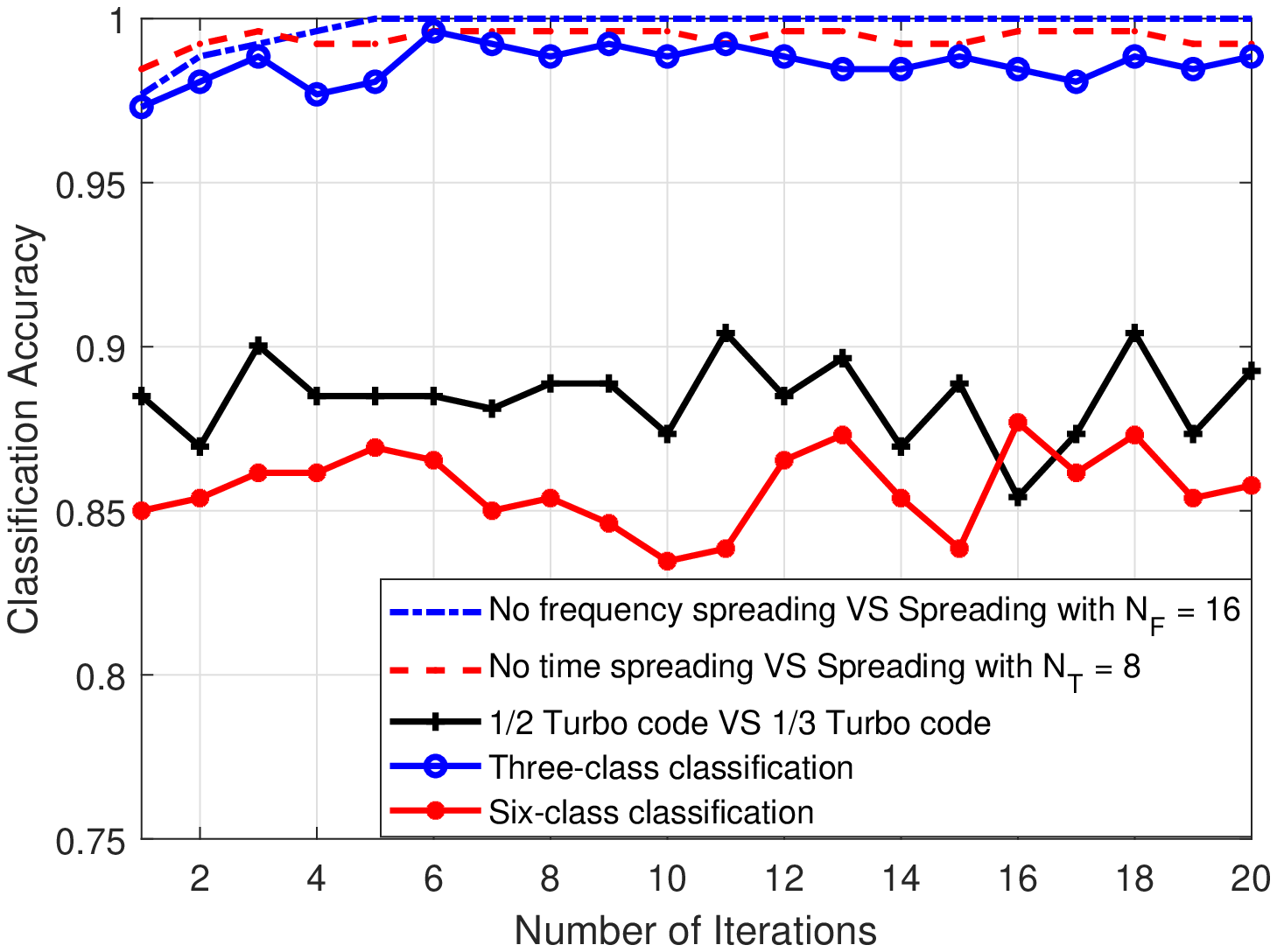}
 \begin{center} 
 (a)
 \end{center}
 \end{minipage}%
 \begin{minipage}[ht]{0.33\linewidth} 
 \centering
 \includegraphics[width=\textwidth]{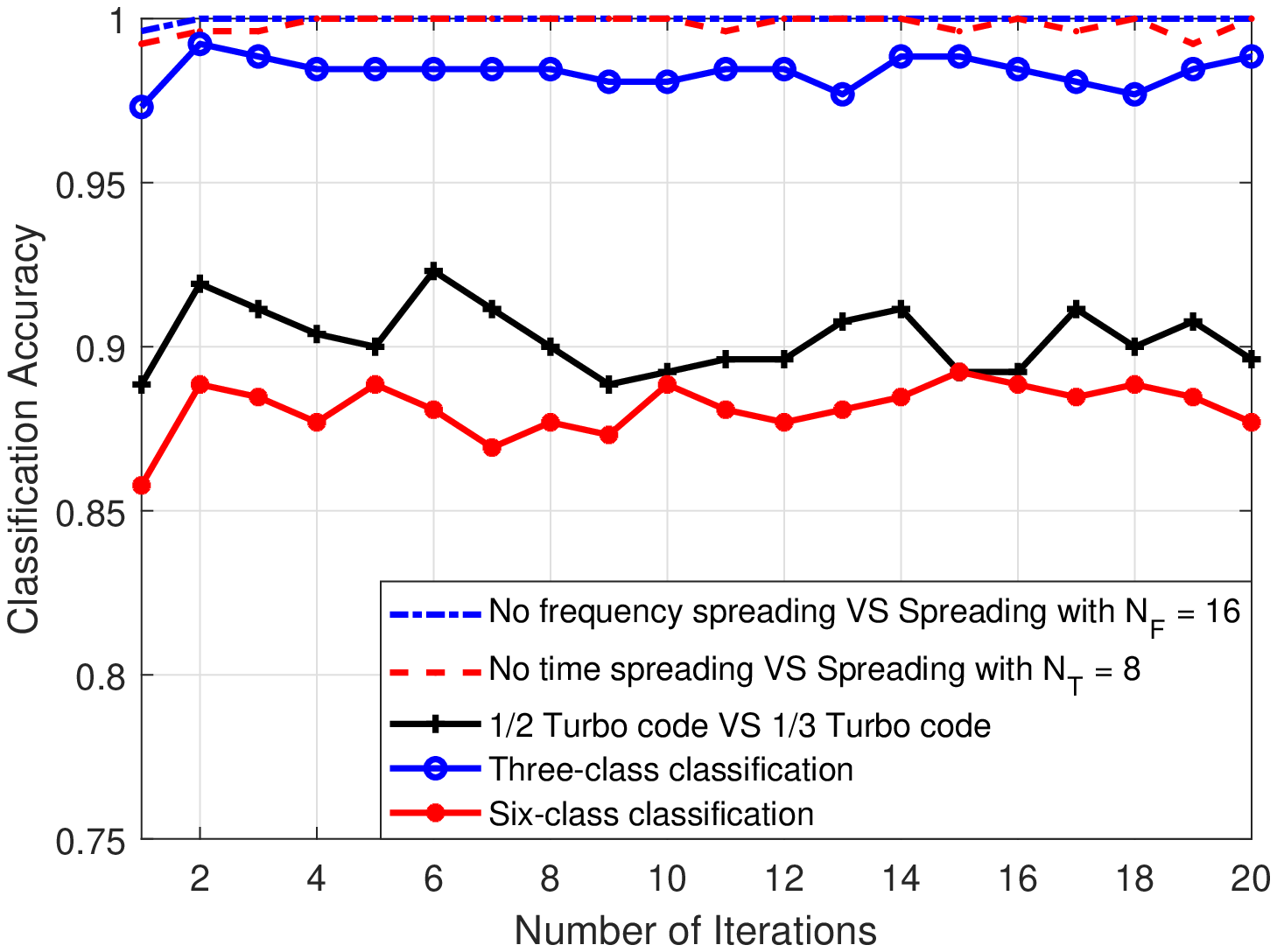}
 \begin{center}
 (b) 
 \end{center} 
 \end{minipage}
 \begin{minipage}[ht]{0.33\linewidth} 
 \centering
 \includegraphics[width=\textwidth]{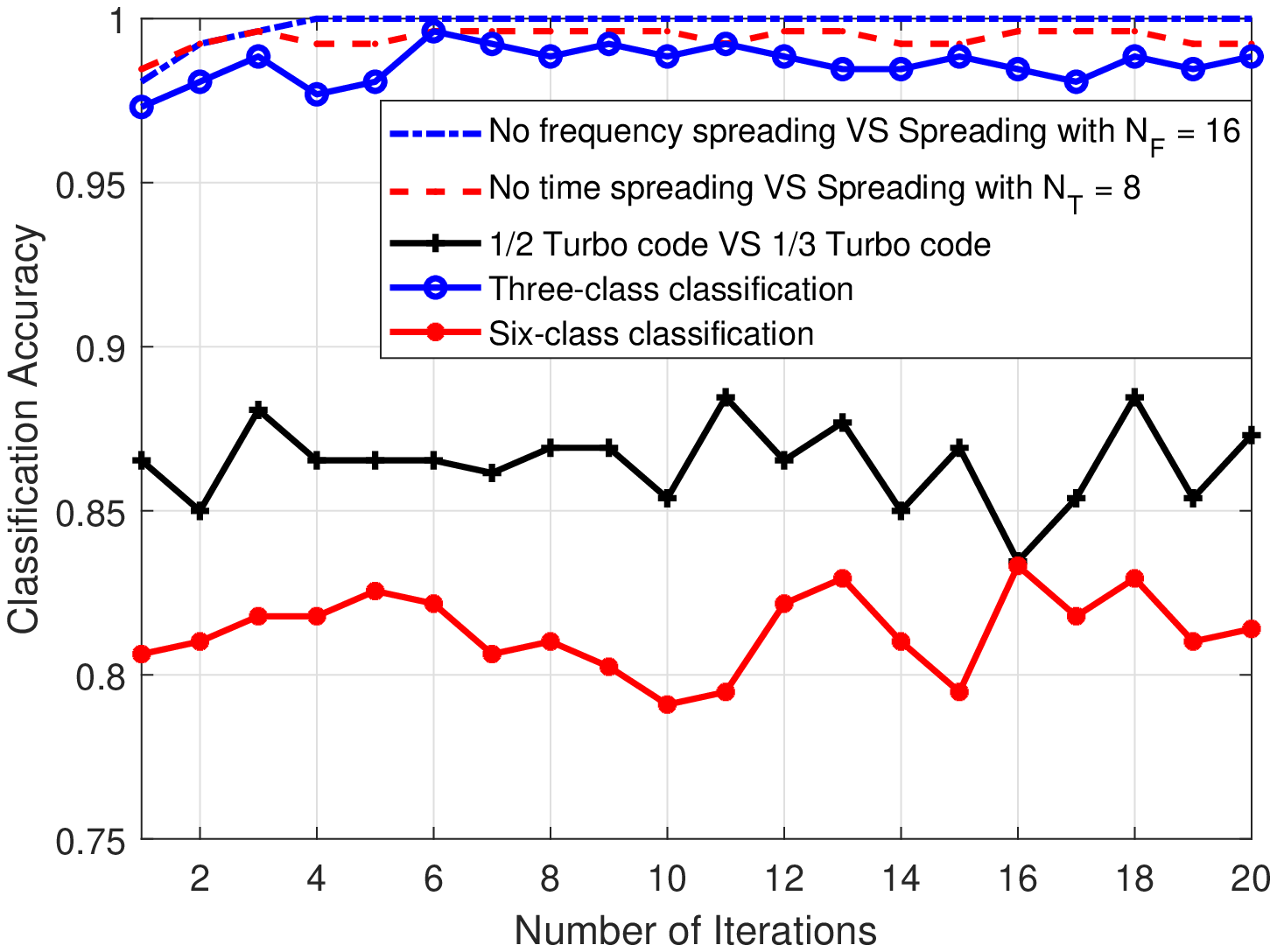}
 \begin{center}
 (c) 
 \end{center} 
 \end{minipage}
 
\caption{Accuracy  of  classifiers  by alternating optimization: (a) LSTM-based RNN classifier; (b) Bi-LSTM classifier; (c)  GRU classifier.
}\label{fig ML classifier alter}
 \end{figure*}

{

In this part, the physical-layer performances of the underwater optical system are presented. Given a modulation and coding rate, the results include the physical-layer frame error rate and throughput v.s. distance for 1-by-100 system without spreading, with spreading $N_F = 16$ and $N_T = 1$, and with spreading $N_F = 16$ and $N_T = 8$.  
From the frame error rate results, it can be observed that the coverage distance is improved for 100 receiving photodetector array and $N_F = 16$ spreading, compared with the non-spreading cases. Then time-domain spreading of $N_T=8$ is introduced to improve performance under the Doppler effect when the AUV speed isn't zero.}
{In Figs.~\ref{fig SIMO TF PHY} and~\ref{fig SIMO TF FER}, it can be observed that the time-domain spreading can achieve improved performance to mitigate the Doppler effect. 

}

The physical-layer throughput is calculated as,
\begin{equation}
 {\tt{Throughput}} = \frac{M N_{TX} N_{SUBC} R_{C}}{2 N_T N_F  T_{OFDM}}  (1-{\tt{FER}}),
\end{equation}
where $N_{SUBC}$ is the number of data subcarriers of one OFDM symbol; $M$ is the order of baseband modulation; $R_{C}$ is the channel coding rate; $\tt{FER}$ is the frame error rate, {which is depicted in Fig.~\ref{fig SIMO TF FER}}; ${T_{OFDM}}$ is the time period of one OFDM symbol. Note that spreading will reduce physical-layer throughput. This is because, in the transmitter there is a constraint that the transmitting power should be a constant value. If multiple users are multiplexed, the per-user power is reduced by the factor of the number of multiplexed users. In the current system design, only one user's signal is spread in time-frequency spreading. {Since only one stream signal is sent at the transmitter without in-phase and conjugate-phone components, the throughput is divided by $2$.}
Note that there are intersection points on the physical-layer-throughput curves in Fig.~\ref{fig SIMO TF PHY}, which are what we utilize to balance the number of photodetectors and the length of the Time-Frequency~(TF) spreading versus distance.

{

The conjugate-symmetric OFDM modulation is adopted in the transmitter so that the generated OFDM symbols are intensity-modulated optical signals without phase modulation. This modulation scheme is especially suited for the optical channel, where the optical scattering can distort the phase coherency. Although there are channel estimation and equalization blocks to recover the phase information, it is a design approach to avoid phase modulation so that the optical signal recovery can be much more robust to the underwater optical scattering effects. By adopting the conjugate-symmetric OFDM modulation, the throughput reduction can be compensated by adjusting the length of the spreading sequence, while the SNR will be increased by combining the conjugate-symmetric constellation symbols at the receiver. The received signal is demodulated with OFDM demodulation, then the symbols and its conjugate parts in the frequency domain are combined before sending the symbols to detection. This operation improves the SNR at the receiver thus improves the coverage of the optical system.

The simulation results in this part will further illustrate what are the coverage and throughput performance with conjugate-symmetric OFDM modulation. In the simulation, the generated optical channel realizations have real-numbered channel coefficients since there is only one stream of the modulated signal without phase modulation. The frame structure with periodic pilot OFDM symbols are adopted at the transmitter, and the channel estimation and symbol equalization are performed at the receiver. The frame error rate and physical-layer throughput results are obtained for a number of simulation setups.
}

{
For $0.1~\rm{m/s}$ AUV speed, where the Doppler shift is $2.10 \times 10^5~\rm{Hz}$, the coverage distance is improved to $23~\rm{m}$ with a throughput of $30~\rm{Mbps}$ by 1-by-100 system with $1/3$ Turbo code, to $30~\rm{m}$ with a throughput of $3~\rm{Mbps}$ by 1-by-100 system with frequency spreading and $1/2$ Turbo code, and to $36~\rm{m}$ with a throughput of $300~\rm{kbps}$ by 1-by-100 system with time-frequency spreading and $1/2$ Turbo code.
For $0.5~\rm{m/s}$ AUV speed, where the Doppler shift is $1.05 \times 10^6~\rm{Hz}$, the coverage distance is improved to $10~\rm{m}$ with a throughput of $30~\rm{Mbps}$ by 1-by-100 system with $1/3$ Turbo code, to $18~\rm{m}$ with a throughput of $3~\rm{Mbps}$ by 1-by-100 system with frequency spreading and $1/2$ Turbo code, and to $24~\rm{m}$ with a throughput of $300~\rm{kbps}$ by 1-by-100 system with time-frequency spreading and $1/2$ Turbo code.}

\section{Machine Learning Results on the Configuration Learning Datasets}\label{sec:learning}

\begin{table*}
\small
\caption[loftitle]{Machine learning performance metrics of the optimal-setting classifiers.}\label{tbl:ML metrics}
\centering
\begin{tabular}{c|c|c|c|c|c|c}\hline
\textbf{Binary Classification} & \textbf{Classifier} & \textbf{Accuracy} &\textbf{Precision} &\textbf{Recall} &\textbf{Specificity} &\textbf{F1 Score}\\
\hline
Binary-class Case~(i) & \emph{SwitchOpt RNN}		& $1.0000  $&$  1.0000  $&$  1.0000  $&$  1.0000  $&$  1.0000$
\\
& LSTM-based RNN		& $1.0000  $&$  1.0000  $&$  1.0000  $&$  1.0000  $&$  1.0000$
\\
& Bi-LSTM &$1.0000  $&$  1.0000  $&$  1.0000  $&$  1.0000  $&$  1.0000$
\\ 
 & GRU		& $1.0000  $&$  1.0000  $&$  1.0000  $&$  1.0000  $&$  1.0000$
\\ 
  & Decision  Tree		&$0.9583  $&$  0.9432 $&$   0.9432  $&$  0.9671  $&$  0.9432$ 
\\
  & Adaptive  Boosting Ensemble		& $0.9792 $&$   0.9545 $&$   0.9882  $&$  0.9742 $&$   0.9711$ \\ 
& SVM &$0.9542  $&$  0.9659   $&$ 0.9140 $&$   0.9796 $&$   0.9392$
\\ \hline
Binary-class Case~(ii) & \emph{SwitchOpt RNN}		& $1.0000  $&$  1.0000  $&$  1.0000  $&$  1.0000  $&$  1.0000$
\\
& LSTM-based RNN & $0.9962  $&$  1.0000   $&$ 0.9925 $&$   1.0000  $&$  0.9962$\\
& Bi-LSTM &$1.0000  $&$  1.0000  $&$  1.0000  $&$  1.0000  $&$  1.0000$
\\
& GRU &$0.9962  $&$  0.9924  $&$  1.0000  $&$  0.9922  $&$  0.9962$
\\
  & Decision  Tree	 & $0.9708  $&$  0.9677  $&$  0.9756  $&$  0.9658   $&$ 0.9717$\\
  & Adaptive  Boosting Ensemble	 & $0.9792 $&$   0.9545 $&$   0.9882  $&$  0.9742  $&$  0.9711$\\ 
& SVM &$0.9583 $&$   0.9677  $&$  0.9524   $&$ 0.9649  $&$  0.9600$
\\ \hline
Binary-class Case~(iii) & \emph{SwitchOpt RNN}		&$0.9231  $&$  0.9032  $&$  0.9655  $&$  0.8696  $&$  0.9333$
\\
& LSTM-based RNN  & $0.9042  $&$  0.9583  $&$  0.8903 $&$   0.9294  $&$  0.9231$
\\
& Bi-LSTM &$0.9231  $&$  0.9032  $&$  0.9655  $&$  0.8696  $&$  0.9333$
\\
& GRU &$0.8846  $&$  0.8103 $&$   0.9792  $&$  0.8036  $&$  0.8868$
\\
& Decision  Tree	  &$0.8792  $&$  0.9028  $&$  0.8966  $&$  0.8526  $&$  0.8997$
\\
& Adaptive  Boosting Ensemble	  & $0.8792  $&$   0.9653  $&$   0.8528  $&$   0.9351  $&$   0.9055$\\ 
& SVM &$0.8792  $&$  0.9444  $&$  0.8662   $&$ 0.9036  $&$  0.9037$
\\ \hline 
Three-Class Classification & \emph{SwitchOpt RNN}	&  $1.0000  $&$  1.0000  $&$  1.0000  $&$  1.0000  $&$  1.0000$
\\
& LSTM-based RNN  & $0.9962  $&$  1.0000   $&$ 0.9925 $&$   1.0000  $&$  0.9962$
\\
& Bi-LSTM & $0.9950 $&$   0.9923  $&$  0.9923  $&$  0.9956 $&$   0.9923$
\\
& GRU		& $1.0000  $&$  1.0000  $&$  1.0000  $&$  1.0000  $&$  1.0000$
\\ 
  & Decision  Tree	  & $0.9478  $&$   0.9083  $&$   0.9033 $&$    0.9657  $&$   0.9058$
\\
  & Adaptive  Boosting Ensemble	  & $0.9088   $&$ 0.8458  $&$  0.8173  $&$  0.9484  $&$  0.8313$\\
& SVM &$0.9499 $&$   0.9125   $&$ 0.9106$&$    0.9608 $&$   0.9115$
\\ \hline
Six-Class Classification & \emph{SwitchOpt RNN}	& $0.8962   $&$   0.8769  $&$  0.8780  $&$  0.9850  $&$  0.8775$
\\
& LSTM-based RNN  &$0.8808  $&$  0.8615   $&$ 0.8462  $&$  0.9777 $&$   0.8538$
\\
& Bi-LSTM & $0.8962   $&$   0.8769  $&$  0.8780  $&$  0.9850  $&$  0.8775$
\\
& GRU & $0.8317  $&$  0.6622 $&$  0.6660 $&$   0.8505  $&$  0.6641$

\\
 & Decision  Tree	  & $0.8084  $&$  0.6229  $&$  0.6543  $&$  0.8441  $&$  0.6382$
\\
& Adaptive  Boosting Ensemble	  & $0.7894  $&$  0.5225 $&$   0.6304  $&$  0.8405  $&$  0.5714$
\\
& SVM &$0.8492  $&$  0.6133  $&$  0.6127  $&$  0.8803 $&$   0.6130$
\\ \hline
\end{tabular}
\end{table*}

{
The configuration learning results are described in this section. The results of the LSTM-based RNN classifier, {Bi-LSTM, GRU}, decision tree, and AdaBoost are illustrated in Fig.~\ref{fig ML classifier} for binary classifications and {multi}-class classifications, respectively. The datasets adopted in the ML algorithms are the signals generated with the physical-layer simulator as well as the coding and spreading configurations maximizing the physical-layer throughput performance. Both binary classification and multi-class classification results are generated. 
}

{
Figs.~\ref{fig ML classifier} depict the 5-fold cross-validation accuracy for binary and multi-class classification by training the pre-detection signal. The binary classifications contain three cases: (i)~``no frequency spreading" VS ``spreading with $N_F = 16$"; (ii)~``no time spreading VS spreading with $N_T = 8$"; (iii)~``$1/2$ Turbo code" VS ``$1/3$ Turbo code".
For the LSTM-based RNN classifier with $600$ hidden units, the classification accuracy of the binary classification cases~(i) and (ii) is $0.99$ and $0.98$ when the numbers of epochs are $4$ and $7$, but decreases when increasing the number of epochs; the accuracy of case~(iii) reaches $0.90$ when the number of epochs is between $28$ and $38$ but decreases due to overfitting with the increment of epochs. The accuracy of case~(iii) is lower than that of cases~(i) and (ii) for two reasons. First, it is indistinct for the bound of the coverage distances of $1/2$ and $1/3$ Turbo code in the short distance. Second, in shorter distances, $1/3$ Turbo code only has a short distance gain but introduces a much lower physical-layer throughput compared with $1/2$ Turbo code, therefore, the classifier is inclined to select the $1/2$ Turbo code in short distances even though $1/3$ Turbo code has a lower frame error rate.
{In contrast, for the Bi-LSTM classifier, the maximal accuracy is $0.99$ for case~(i), $0.98$ for case~(ii) and $0.90$ for case~(iii).
For the GRU classifier, the maximal accuracy is $1$ for case~(i), $0.96$ for case~(ii) and $0.86$ for case~(iii).}
For the decision tree, the binary classification accuracy reaches $0.94$ for case~(i), $0.98$ for case~(ii) and $0.86$ for case~(iii), which is lower than that of LSTM-based RNN.
Fig.~\ref{fig ML classifier}(e) depicts the accuracy of AdaBoost aggregation method, where we can see that for cases~(i) and (ii), the accuracy converges to $0.98$ and $0.97$ when the number of learning cycle is greater than $30$, and for case~(iii), the accuracy has an upper bound of $0.85$. 
{Among the ML classifiers, the GRU classifier performs the best for binary case~(i) in classification acccuracy performance. The LSTM-based RNN classifier and Bi-LSTM classifier perform the best for binary cases~(ii) and~(iii) within the range of non-overfitting.}

The classes of three-class classification contain: (1) ``No spreading", (2) ``Spreading with $N_F = 16$, $N_T = 1$, (3) ``Spreading with $N_F = 16$, $N_T = 8$. The classes of six-class classification contain: (1) ``No spreading, $1/2$ Turbo code", (2) ``No spreading, $1/3$ Turbo code", (3) ``Spreading with $N_F = 16$, $N_T = 1$, $1/2$ Turbo code", (4) ``Spreading with $N_F = 16$, $N_T = 1$, $1/3$ Turbo code", (5) ``Spreading with $N_F = 16$, $N_T = 8$, $1/2$ Turbo code", (6) ``Spreading with $N_F = 16$, $N_T = 8$, $1/3$ Turbo code".}

For the LSTM-based RNN classifier, the accuracy of three-class and six-class classification reaches $0.96$ and $0.83$ at the point when the number of epochs is $9$ and $13$, but decreases due to overfitting with more epochs.
{
For the Bi-LSTM classifier, the accuracy for three-class is $0.96$, close to that of LSTM-based RNN classifier, and the accuracy for six-class is $0.84$. For GRU, the accuracy is lower than $0.93$ for three-class and is lower than $0.82$ for six-class.}
In contrast, for the decision tree, it has a lower accuracy than LSTM, where the accuracy converges to $0.92$ for three-class classification, to $0.80$ for six-class classification. The AdaBoost has an upper bound around $0.85$ for three-class classification and around $0.69$ for six-class classification. Therefore, we can conclude that the LSTM-based RNN classifier performs best among the ML classifiers evaluated for multi-class classification with its optimal setting.

{
{The configuration learning problem in this underwater optimal system is to identify the optimal ML classifier that maximizes the classification accuracy performance. The optimal ML classifier and its parameters depend on the signal characteristics. In particular, for the binary classification case~(iii), the ``$1/2$ Turbo code" VS ``$1/3$ Turbo code" case, the Bi-LSTM achieves the best performance with the tunable parameter of the number of hidden units. This parameter represents the information stored in Bi-LSTM between the time steps. While for the binary classification case~(i), ``no frequency spreading" v.s.``spreading with $N_F = 16$", the Bi-LSTM performs close to the LSTM. For the binary classification case~(ii), ``no time spreading VS spreading with $N_T = 8$" Bi-LSTM also has the close performance to LSTM. The reasons of these performance results depend on the how the ML classifiers responds to different signal characteristics, including the signals with varied spreading schemes and signals with varied coding rates. It is necessary to further evaluate the tuning of the number of hidden units of the Bi-LSTM for the three binary classification cases to observe how to optimize classification accuracy performance with the optimization variables of the number of hidden units. The evaluations of the LSTM, Bi-LSTM, and GRU are done for the three binary cases and the two multi-class classifications. The results are plotted in Fig.~\ref{fig ML classifier alter}(a),~(b) and~(c). The range of number of hidden units is $200 \sim 4000$, and the range of the epochs is $5 \sim 50$. Performance upper bounding effects with the increasing of the number of units are observed for the classification accuracy results in LSTM, Bi-LSTM, and GRU for all the binary and multi-class cases. The classification accuracy performance results of three classifiers are the highest for the binary classification case~(i), the configuration learning of the waveforms of switching on/off the frequency spreading, compared with other binary cases of switching on/off time spreading and varying the Turbo coding rate. These results indicate that the classification of the frequency spreading configuration has the highest accuracy, therefore the frequency spreading configuration should be adjusted before adjusting the time spreading and the coding rate, based on the proposed configuration learning framework. In terms of the multi-class cases, the three-class classification has a high level of accuracy for LSTM, indicating that the proposed configuration learning framework can effectively select the optimal configurations based on the LSTM classifier. 

In Table~\ref{tbl:ML metrics}, the classification metrics including accuracy, precision, recall, specificity and F1 Score are calculated for the binary and multi-class classification with optimal-setting classifiers. {For all these metrics, it can be observed that, for binary case~(i), the LSTM-based RNN, Bi-LSTM and GRU classifier perform the best among the ML classifiers with accuracy of $1$. For  binary case~(ii) and (iii) and six-class classifications, the Bi-LSTM performs the best. For three-class classification, the GRU performs the best. The other competing classifiers do not show high classification accuracy performance in general for the binary and multi-class classification in this configuration learning problem. The \emph{SwitchOpt RNN} outperforms all the classifiers by selecting the optimal ML classifiers based on observing the performances of the candidate classifiers.}

}



}

\section{Conclusion}\label{sec:conc}
{
Our proposal designs the configuration learning for an underwater wireless optical communication system. The configuration learning problem is introduced by optimizing the Machine Learning~(ML) classifiers to achieve the classification accuracy performance of the transmitter configuration. To train the ML classifier, the received signal waveform and the optimal transmitter configuration maximizing the physical-layer throughput are obtained from a bit-level physical-layer simulator. The ML classifiers evaluated include the LSTM-based RNN, Bi-LSTM, GRU, decision tree, AdaBoost, and SVM classifiers. An algorithm based on alternating optimization and switching of ML classifier is proposed. The results indicate that, the proposed algorithm outperforms competing algorithms. Our configuration learning framework and the designed algorithm can be applied to next-generation cellular networks and signal processing systems. Our future work will consider more modulation schemes that are robust to the dispersion of the underwater optical channel.
}

\bibliographystyle{IEEEtran}
\bibliography{reference_uwcellular}

{
%
}

\end{document}